\documentclass[ aps,
    https://www.overleaf.com/project/5eb9202193f026000132e2d7            pra,
                twocolumn,
                longbibliography,
                superscriptaddress
                ]{revtex4-1}
\usepackage{graphicx}
\usepackage[normalem]{ulem}
\usepackage[dvipsnames]{xcolor}
\usepackage{amsmath}
\usepackage{amssymb}
\usepackage{braket}
\usepackage[colorlinks=true,urlcolor=blue,citecolor=blue,linkcolor=magenta]{hyperref}

\begin{document}
\title{Identifying nonclassicality from experimental data using artificial neural networks}
\author{Valentin Gebhart}
\email{gebhart@lens.unifi.it}
\affiliation{QSTAR, INO-CNR, and LENS, Largo Enrico Fermi 2, I-50125 Firenze, Italy}
\affiliation{Universit\`a degli Studi di Napoli ”Federico II”, Via Cinthia 21, I-80126 Napoli, Italy}

\author{Martin Bohmann}
\email{martin.bohmann@oeaw.ac.at}
\affiliation{Institute for Quantum Optics and Quantum Information - IQOQI Vienna, Austrian Academy of Sciences, Boltzmanngasse 3, 1090 Vienna, Austria}

\author{Karsten Weiher}
\affiliation{Institut f\"ur Physik, Universit\"at Rostock, D-18051 Rostock, Germany}

\author{Nicola Biagi}
\affiliation{Istituto Nazionale di Ottica (CNR-INO), L.go E. Fermi 6, 50125 Florence, Italy}
\affiliation{LENS and Department of Physics $\&$ Astronomy, University of Firenze, 50019 Sesto Fiorentino, Florence, Italy}

\author{Alessandro Zavatta}
\affiliation{Istituto Nazionale di Ottica (CNR-INO), L.go E. Fermi 6, 50125 Florence, Italy}
\affiliation{LENS and Department of Physics $\&$ Astronomy, University of Firenze, 50019 Sesto Fiorentino, Florence, Italy}

\author{Marco Bellini}
\affiliation{Istituto Nazionale di Ottica (CNR-INO), L.go E. Fermi 6, 50125 Florence, Italy}
\affiliation{LENS and Department of Physics $\&$ Astronomy, University of Firenze, 50019 Sesto Fiorentino, Florence, Italy}

\author{Elizabeth Agudelo}
\email{elizabeth.agudelo@oeaw.ac.at}
\affiliation{Institute for Quantum Optics and Quantum Information - IQOQI Vienna, Austrian Academy of Sciences, Boltzmanngasse 3, 1090 Vienna, Austria}

\begin{abstract}
    The fast and accessible verification of nonclassical resources is an indispensable step toward a broad utilization of continuous-variable quantum technologies. 
    Here, we use machine learning methods for the identification of nonclassicality of quantum states of light by processing experimental data obtained via homodyne detection. 
    For this purpose, we train an artificial neural network to classify classical and nonclassical states from their quadrature-measurement distributions. 
    We demonstrate that the network is able to correctly identify classical and nonclassical features from real experimental quadrature data for different states of light. 
    Furthermore, we show that nonclassicality of some states that were not used in the training phase is also recognized.
    Circumventing the requirement of the large sample sizes needed to perform homodyne tomography, our approach presents a promising alternative for the identification of nonclassicality for small sample sizes, indicating applicability for fast sorting or direct monitoring of experimental data. 
\end{abstract}
\maketitle

\section{Introduction}	
    
    Quantum technologies promise various advantages over classical technologies. 
    By employing different features of quantum systems that are not present in classical systems, one can, e.g., perform more precise measurements, speed up computations, or share information in a more secure way. 
    These nonclassical properties create possibilities to optimally exploit physical systems for many technological challenges. 
    Light fields, described as continuous-variable systems, play a key role for the transmission and manipulation of quantum information \cite{braunstein2005}. 
    Due to their infinite dimensions and an accessible control by means of linear optical elements and homodyne detection, they are widely considered for quantum technological applications. 
    In the case of single-mode continuous-variable quantum systems, the central quantum resource is nonclassicality \cite{streltsov_2017,sperling_2018}.
    Directly related to the negativities \cite{titulaer_1965,mandel_1986} of the Glauber-Sudarshan $P$ representation of the quantum state \cite{glauber_1963,sudarshan_1963}, nonclassicality manifests itself in different observable characteristics such as photon antibunching \cite{Carmichael_1976,Kimble_1976,kimble_1977}, sub-Poissonian photon-number statistics \cite{mandel_1979,Zou_1990}, and quadrature squeezing \cite{Yuen_1976,Walls_1983,Caves_1985, Loudon_1987,Dodonov_2002}, and can be transformed into other quantum resources such as entanglement \cite{vogel_2014,killoran_2016}.
    The fundamental nature of nonclassicality is exploited for the investigation of the roots of quantum phenomena and several quantum technological tasks such as, e.g., precision measurements.

    Due to its crucial importance for quantum technologies, a fast and reliable identification of nonclassicality from experimental observations of the quantum state represents an unavoidable step toward a practical usage of such a resource for quantum technologies. 
    In continuous-variable systems, one of the most common measurement methods is homodyne detection \cite{welsch_1999}.
    Advanced state tomography techniques based on this type of measurements have been developed \cite{paris_2004,lvovsky_2009}. 
    However, nonclassicality certification based on homodyne tomography usually requires many different quadrature measurements and involved analysis tools.
    A different approach is nonclassicality certification via negativities of reconstructed quasiprobabilities \cite{cahill_1969b} (particularly, the Glauber-Sudarshan P function \cite{Kiesel_2008} and the Wigner function \cite{smithey_1993, Dunn_1995, leibfried_1996, deleglise_2008}).
    Methods that involve regularizations of quasiprobabilities have been implemented for the single-mode and multimode scenarios \cite{kiesel_2010, agudelo_2013}, and more recently, phase-space inequalities have been proposed and tested experimentally \cite{bohmann_2020, bohmann_2020b, biagi_2020}. 
    Finally, a direct nonclassicality estimation without the need for quantum state tomography was proposed in Ref. \cite{mari_2011}. Here, the nonclassicality of phase randomized states was classified via semidefinite programming.
    In all above approaches, to guarantee the detection of nonclassicality with a high statistical significance, extensive measurements must be performed (using different measurement settings or sampling different moments), after which advanced postprocessing is required (estimation of pattern functions, reconstruction of quasiprobabilities, and semidefinite programming, among others).
    Consequently, these methods are often complex and time consuming.
    A direct access to nonclassicality identifiers from unprocessed and finite homodyne-detection data is therefore desirable.
    
    In recent years, the problem of the classification of unstructured and complex data has been increasingly addressed with the help of machine learning (ML) techniques \cite{nielsen_2015}.
    In the quantum domain, a wide range of challenges was tackled using various different forms of ML, see, e.g., \cite{hentschel_2010,wiebe_2014,magesan_2015,krenn_2016,van_nieuwenburg_2017,carleo_2017,torlai_2018,lumino_2018,bukov_2018,fosel_2018,canabarro_2019,cimini_2019,nautrup_2019,agresti_2019,gebhart_2020,cimini_2020,tiunov_2020,nolan_2020,dunjko_2018,ahmed_2020} for a review. 
    ML tools have been applied to the identification of nonclassicality \cite{cimini_2020,gebhart_2020}. 
    In Ref. \cite{gebhart_2020}, neural networks (NNs) were trained to identify nonclassicality from simulated data of multiplexed click-counting detection schemes, and in Ref. \cite{cimini_2020}, the networks were trained to detect the negativity of the multimode Wigner function using results from multimode homodyne detection measurements. 
    Also, ML in the form of restricted Boltzmann machines have been used to perform homodyne tomography in Ref. \cite{tiunov_2020}.

    

    In this paper, we use ML techniques to identify nonclassicality of single-mode states based on a finite number of quadrature measurements recorded via balanced homodyne detection. 
    For this purpose, we employ a dense artificial NN and train it with supervised learning of simulated homodyne detection data from several noisy classical and nonclassical states.
    We demonstrate the successful performance of the NN nonclassicality prediction on real experimental data and compare the results with established nonclassicality identification methods.
    Furthermore, we test the performance of the network for experimentally generated states which were not used in the training procedure and show that the NN can identify different nonclassical features at once. 
    We conclude that the ML approach offers an accessible alternative for the classification of single-mode nonclassicality, and, particularly, due to its performance on small sample sizes, the presented approach constitutes a powerful tool for data pre-selecting, sorting, and on-site real-time monitoring of experiments.
    Our result represents an approach to train NNss for identifying nonclassicality of single-mode phase-sensitive states, here measured by homodyne detection.
    
    The paper is structured as follows.
    In Sec. \ref{sec:2}, we briefly recall the technique of single-mode balanced homodyne detection.
    In Sec. \ref{sec:3}, we describe in detail the training of the NN and the resulting nonclassicality identifier.
    In Sec. \ref{sec:4}, we apply the NN to experimental homodyne measurement data and then analyze its performance on untrained data in Sec. \ref{sec:5}.
    We summarize and conclude in Sec. \ref{sec:6}.

\section{Balanced homodyne measurement and nonclassical states}\label{sec:2}

\begin{figure}[t]
		\center
		\includegraphics[width = 0.5\columnwidth]{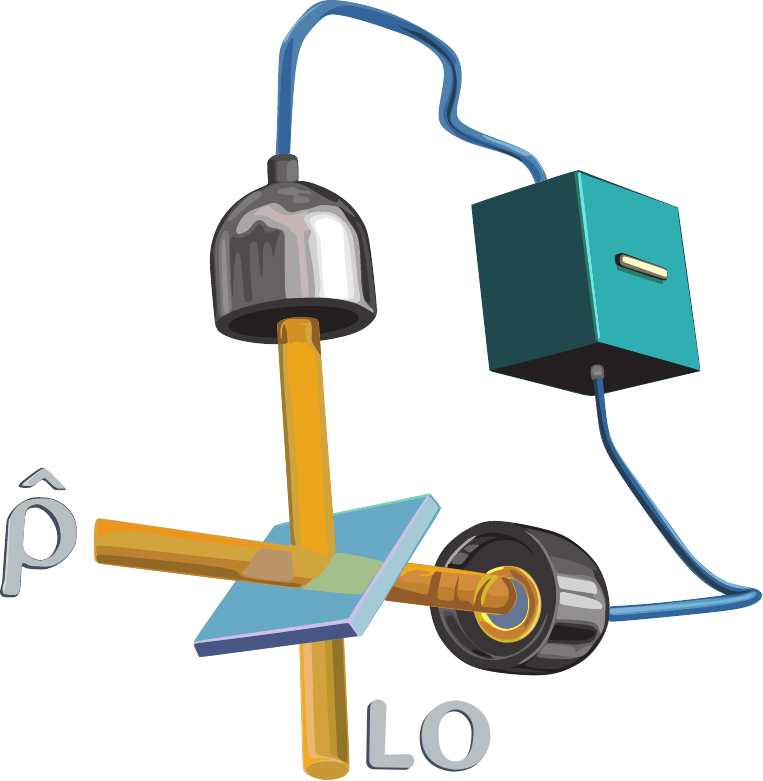}
		\caption{
		    Homodyne detection scheme. 
		    The signal field $\hat\rho$ and the reference coherent beam (LO) are mixed using a 50:50 beam splitter (BS) before the measuring light intensity.
		}
		\label{fig:bhd}
	\end{figure}

    Any direct experimental investigation of light is based on photodetection.
    Depending on the information on the quantum statistics of the measured light required, different measurement schemes need to be implemented.
	For example, photon-counting measurements are not sensitive to the phase of the sensed field.
	To get 
	information about the phase, interferometric methods have to be applied.
	In these methods, the field is mixed with a reference beam, the so-called local oscillator (LO).
	The mixing takes place just before intensity measurements~\cite{welsch_1999, lvovsky_2009}.
	The scheme of balanced homodyne detection is shown in Fig. \ref{fig:bhd}.
	It consists of the signal field $\hat\rho$, the LO, a 50:50 beam splitter (BS), two proportional photodetectors, and the electronics used to subtract and amplify the photocurrents after all. 
    Homodyning with an intense coherent LO gives the phase sensitivity necessary to measure the quadrature variances
    \cite{Carmichael_1987,Braunstein_1990, Vogel_1993}.
	
	This kind of interferometric approach is necessary for the reconstruction of the quasiprobabilities of bosonic states.
	In principle, all normally ordered moments can be determined from this measurement scheme, including the ones which contain different numbers of creation and annihilation operators.
	Thus, homodyne detection drastically enlarges our measuring capabilities in a simple way. 
	The key for the quasiprobabilitiy estimation is to perform measurements for a large set of quadrature phases, which leads ultimately to a proper state reconstruction.
	Balanced homodyne detection and the subsequent reconstruction of the Wigner function have become a standard measuring technique in quantum systems such as, e.g., quantum light, molecules, and trapped atoms \cite{smithey_1993, Dunn_1995, leibfried_1996, deleglise_2008}. 
	
    Although experimentally accessible, phase-space function reconstructions and moment-based nonclassicality criteria require significant amounts of measurement data, computational power, and postprocessing time. 
    Here, we propose a shortcut to this process.
    Using NNs, we can do an on-the-fly nonclassicality identification with few measurements.


\section{Training the NN}\label{sec:3}

\subsection{Setup of the network}
    The input vector of the network consists of a normalized histogram (relative frequencies) of homodyne-detection data which is collected along a fixed phase setting. 
    To generate the histogram from simulated or experimentally generated data (produced from quadrature-measurement outcomes $x$), we bin the data into $160$ equally sized intervals which cover the interval $[ -8,8]$ \cite{Note1}.
    Since the histogram is normalized, input vectors constructed from arbitrary numbers of detection events can be used for the same network.
    
    We use a fully connected artificial NN with an input layer of size $160$, an output layer of size $2$ and three hidden layers with sizes $64$, $32$ and $16$. 
    The hidden layers are activated with the rectified linear unit, and the output layer is activated with a softmax function. 
    These parameters were chosen for a good performance in discriminating between classical and nonclassical states.
    The simulated data consisting of $2\times 10^4$ input vectors per training family (see below) are split into training data ($80\%$) and validation data ($20\%$). The network is trained until the validation error stops decreasing for more than $10$ training cycles.

    Considering the experimental data on which we want to test the network's prediction later, we simulate $16000$ detection events to generate each training input vector.
    We train the NN with data generated from Fock, squeezed-coherent, and single-photon-added coherent states (SPACS) as states that show nonclassical signatures and with coherent, thermal, and mixtures of coherent states as states showing classical characteristics, see Appendix \ref{sec:appendix_traingdata} for a discussion of this choice.
    All families of states used in the training are summarized together with their parameters in Appendix \ref{sec:Appendix}.
    To account for realistic (imperfect) scenarios, we chose an overall efficiency of the homodyne measurement of $\eta=0.6$ \cite{biagi_2020}.
    Note that the quantum efficiency, that represents external limitations such as channel or detector efficiencies, can equivalently be used to describe noisy quantum states.
    Thus, we train the network with data that correspond to the detection of realistic, lossy quantum states.

\subsection{Identification of nonclassicality} 
    
    In the training process, we assign the value $0$ to all classical quadrature data and the value $1$ to nonclassical data.
    The output of the NN is a value $r$ between $0$ and $1$ that provides a way to discriminate classical and nonclassical data.
    A high output value (close to $1$) indicates the nonclassical character of the tested quadrature data.
    We choose a threshold value $t$ above which we say that the NN identifies nonclassicality.
    As our goal is to faithfully identify nonclassicality, we set $t=0.9$.
    This means that, for $r>t=0.9$, we conclude that the NN identifies nonclassicality.
    In this way, we might reject some nonclassical states to be recognized as such, but we minimize the risk of falsely recognizing classical states as nonclassical ones.
    Note that depending on the specific requirements and the choice of trained and studied states, the value of $t$ can be adapted.
    
    In this context, it is important to stress that the result of the NN can only be an indication for nonclassical states; cf. also Ref. \cite{gebhart_2020}.
    A certification of nonclassicality requires full analysis including the evaluation of a nonclassicality test (witness) and a proper treatment of errors.
    While such an analysis can be rather involved, the proposed NN approach allows one to implement an easy and fast identification of nonclassicality.
    Therefore, it provides a useful tool for pre-selecting and sorting of data or the online, in-laboratory monitoring of experiments.

\subsection{Performance of the network on trained states}

\begin{figure}[t]
		\center
		\includegraphics[width = 0.9\columnwidth]{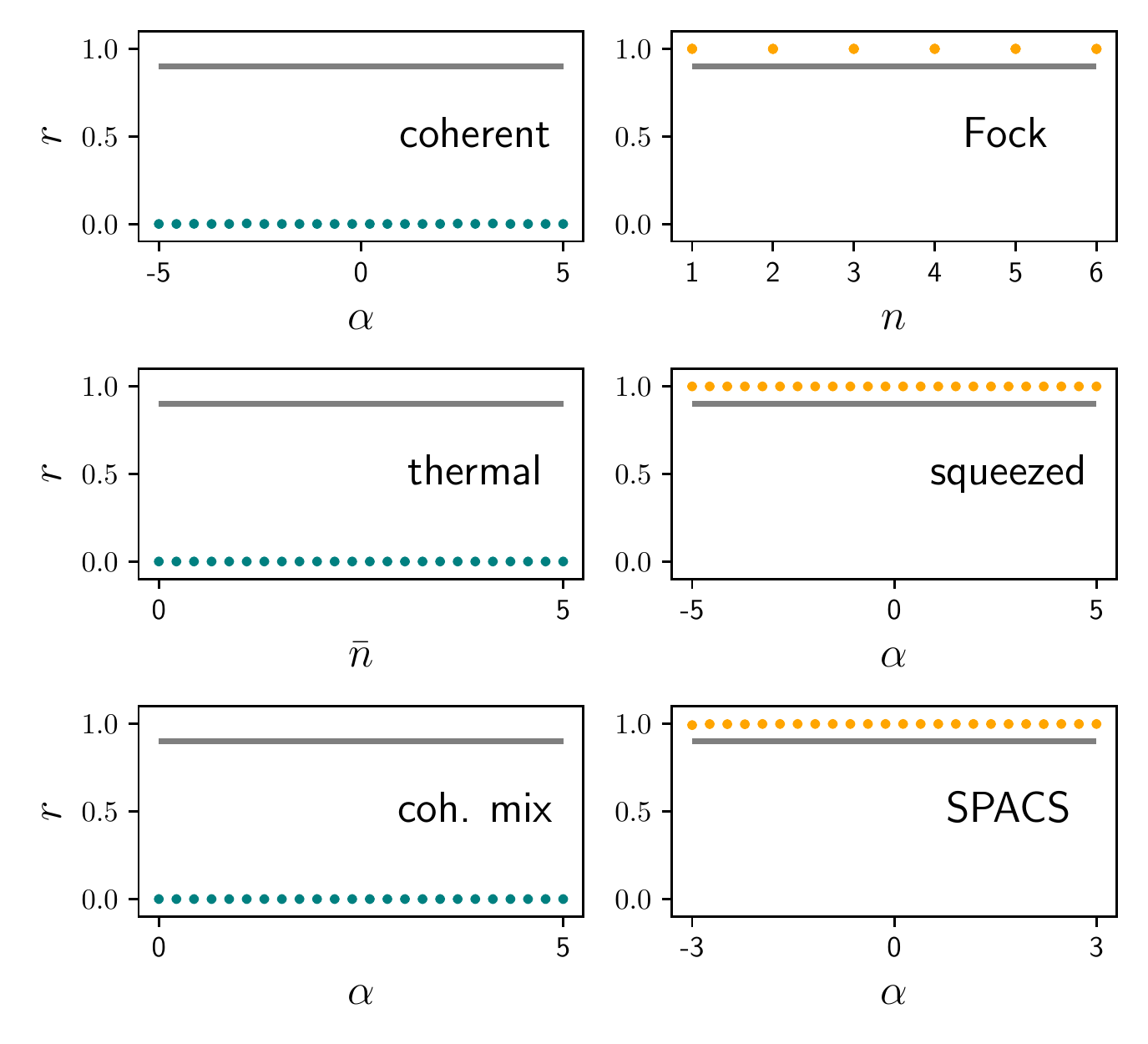}
		\caption{
		    Nonclassicality prediction of the neural network (NN) on the training states [coherent, thermal and mixed coherent states as classical ones; Fock, squeezed-coherent and single-photon-added coherent states (SPACS) as nonclassical ones], each in its corresponding state-parameter domain.
		    $\alpha$ is the coherent amplitude, $n$ is the number of photons, and $\bar n$ is the mean number of photons. 
		    The gray horizontal line corresponds to the nonclassicality threshold $t=0.9$.
		    Note that for the squeezed-coherent states, the squeezing parameter $\xi$ is chosen randomly in $\xi\in[0.5,1]$ and is not shown in this plot. For each Fock state, the NN prediction is tested for four different simulations of the quadrature measurements.
		    For details on the state parameters, see Appendix \ref{sec:Appendix}. 
		}
		\label{fig:performance_training}
	\end{figure}

    In Fig. \ref{fig:performance_training}, we show the output $r$ of the network for the different families of training states in their corresponding parameter ranges. 
    All training families are correctly and consistently recognized to be classical or nonclassical.
    This holds for the total parameter regions of the considered states (cf. Appendix \ref{sec:Appendix}), indicating that the training of the NN is successful in the sense that the network learned to correctly classify the states from the training set into classical and nonclassical ones.

\section{Application to experimental data}\label{sec:4}
    
    Here, we will use the trained NN for the identification of nonclassicality from experimental quadrature data.
    We analyze data from two different families of states: single-mode squeezed states and SPACS.
    This analysis will demonstrate the strength of the network approach as a fast and easy-to-implement characterization tool for experimental data.

\subsection{Squeezed vacuum states}
    The first nonclassical experimental state we consider is a squeezed vacuum state. 
    The vacuum state is squeezed along the real axis of the coherent plane. 
    Details on the experimental realization can be found in Ref.~\cite{agudelo_2015}.
    
    In the measurements, the  homodyne phase setting is changed continuously within the interval $\phi\in[0,2\pi]$. 
    The resulting measurement data are then divided into $125$ bins of size $\Delta\phi=2\pi/125$, such that $\sim 16000$ detection events are grouped together to constitute an input vector of the NN.   
    For our analysis, the amount of squeezing $|\xi_\mathrm{exp}|$ and the quantum efficiency $\eta_\mathrm{exp}$ of the detectors do not have to be known, which highlights the practicability of the NN prediction.
    
\begin{figure}[t]
		\center
		\includegraphics[width=0.9\columnwidth]{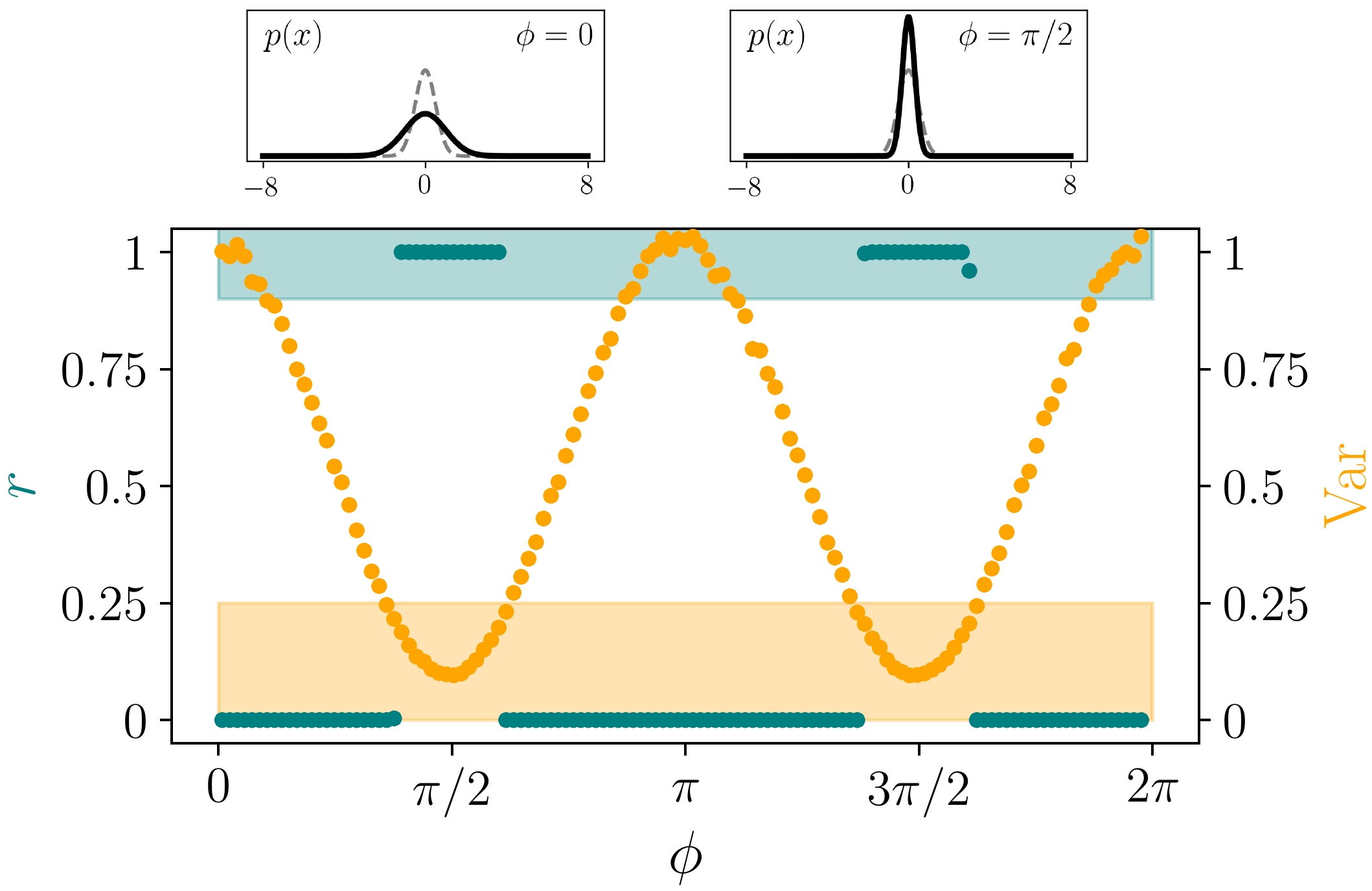}
		\caption{
            Bottom: Nonclassicality prediction of the neural network (NN; teal) and quadrature variance of the corresponding distribution (yellow) for experimental data from squeezed states, dependent on the homodyne phase setting. Shaded regions are indicating nonclassicality for the corresponding quantity. Top: For $\phi=0$ and $\phi=\pi/2$, the exact quadrature distribution $p(x)$ is shown (solid), in comparison with the quadrature distribution of the vacuum state (dashed). 
		}
		\label{fig:squeezed}
    \end{figure}

    In Fig. \ref{fig:squeezed} (bottom), we show the prediction of the network for the nonclassicality of the squeezed state with respect to the homodyne phase setting together with the variance of the measured quadrature distribution. 
    Additionally, the quadrature distributions $p(x)$ for $\phi=0$ and $\phi=\pi/2$ (solid) compared with the vacuum quadrature distribution (dashed) are displayed (top).
    It is known that nonclassicality in quadrature data can be verified by observing single-mode quadrature squeezing, see, e.g., Ref. \cite{vogel_2006}.
    That is, if the quadrature variance $\mathrm{Var[\hat x(\phi)]}$ is below the vacuum noise for some values of $\phi$, $\mathrm{Var[\hat x(\phi)]}<1/4$, nonclassicality is detected.
    We see that the domain of nonclassicality classification of the network coincides well with the domain of nonclassicality detection by sub-shot-noise variance.
    
    In short, we confirm that the NN learns the standard nonclassicality classifier of sub-shot-noise variance.
    If one is simply interested in the detection of squeezing, measuring the variance of the quadrature distribution remains sufficient.
    However, as discussed below, in contrast to a mere variance classifier, the NN can learn how to identify further nonclassicality features.
    It is more flexible than the squeezing condition which recognizes only one specific nonclassical feature, and it can be advantageous in scenarios where the underlying quantum state is not known and cannot be captured by a simple variance condition.

\subsection{SPACS}

    Let us now analyze the prediction of the network for experimentally generated SPACS, which are the result of the single application of the photon creation operator onto a coherent state. In principle, such states are always nonclassical, independent of the input coherent state; however, they present an evident Wigner negativity and resemble single-photon Fock states only for small coherent state amplitudes. On the other hand, for intermediate amplitudes, they also present quadrature squeezing. Exhibiting a variety of different quantum features in different parameter regions, SPACS are therefore particularly interesting candidates for testing the performance of the NN.
    The experimental data consist of quadrature values, measured via homodyne detection, for the states $\mathcal{N}\hat{a}^\dagger \left | \alpha\right\rangle$ ($\mathcal{N}$ is a normalization constant) with $14$ different values of $\alpha\in \mathbb{R}^+$. To experimentally generate such optical states, we injected the signal mode of a parametric down conversion crystal with coherent states obtained from the 786 nm emission of a Ti:Sa mode-locked laser \cite{zavatta_2005}.
    When the same crystal is pumped with an intense ultraviolet beam, obtained by frequency doubling the same laser, the detection of an idler photon heralds the addition of a single photon onto the seed coherent state. In other words, each idler detection event announces the presence of SPACS along the signal mode. Performing heralded homodyne detection on this mode, we measured the quadrature distributions along $11$ different quadrature angles $\phi$ for each value of $\alpha$ \cite{zavatta_2005}. Mode mismatch between the seed coherent states and the pump and LO beams, optical losses, electronic noise, and limited detector quantum efficiency in the homodyne measurement setup are the main causes for a non-unit overall efficiency of $\eta_{\mathrm{exp}}\approx0.6$ in the experiment.
    For each state, $15963$ detection events are used to construct the network input vector. 
    
\begin{figure}[t]
		\center
		\includegraphics[width = 0.9\columnwidth]{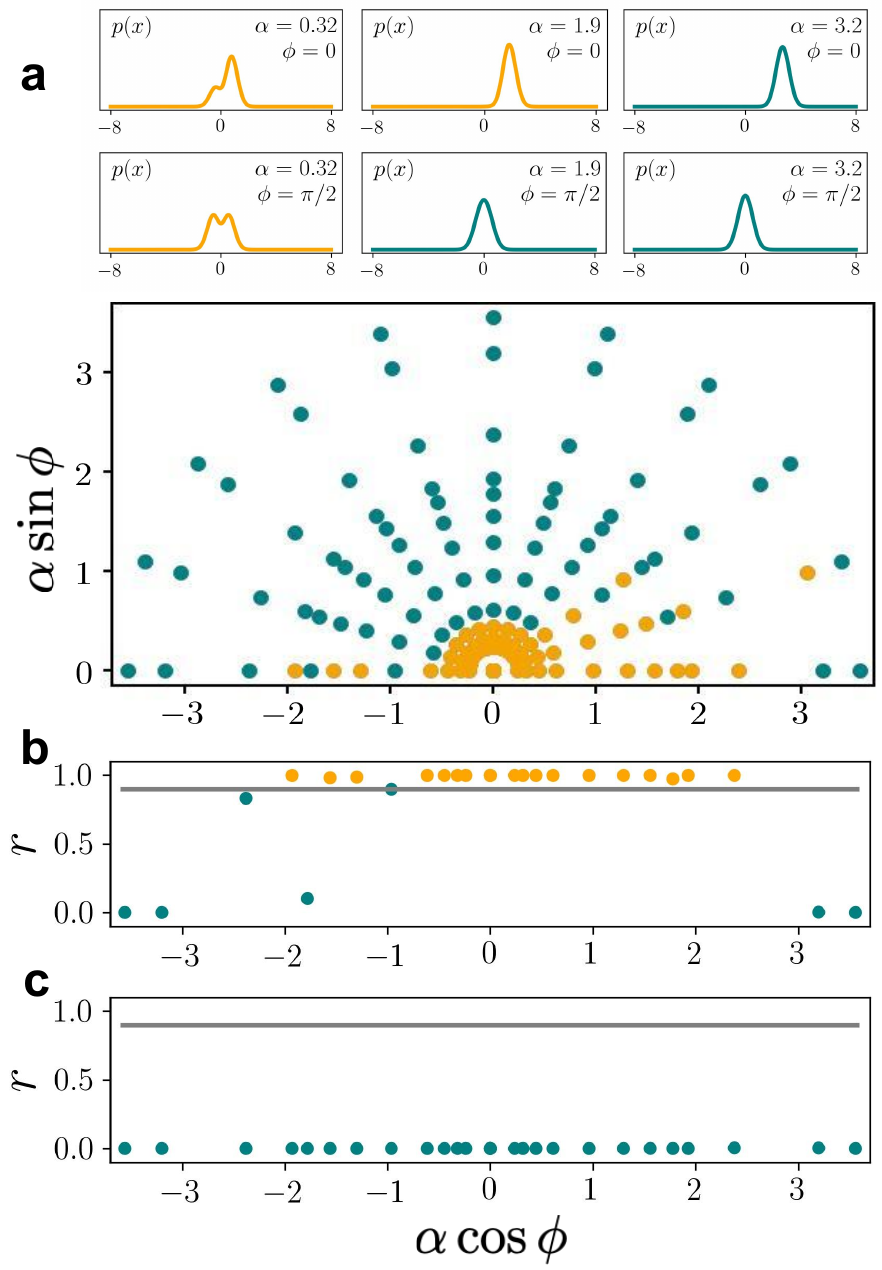}
		\caption{
		    (a) (bottom) Binary neural network (NN) prediction for experimentally generated single-photon-added coherent states (SPACS; $\hat{a}^\dagger \left |\alpha\right\rangle$ with $14$ different values of $\alpha$ measured along $11$ quadrature angles $\phi$). Yellow (teal) points indicate that the prediction $r$ of the NN is above (below) the threshold $t=0.9$.
		    (top) Additionally, we exemplarily display the quadrature distribution $p(x)$ for some representative $\alpha$ and $\phi$.
		    (b) Cut of the network prediction for SPACS along $\sin\phi=0$.
		    (c) Prediction of the NN for experimental data from coherent states for the same parameters as in (b).
		}
		\label{fig:performance_spacs}
	\end{figure}

    In Fig. \ref{fig:performance_spacs}(a), we show the (binary) prediction of the network for the experimental SPACS data, together with exemplary quadrature distributions $p(x)$ for different combinations of $\alpha$ and $\phi$.
    We observe that the ability of the NN to identify nonclassicality depends crucially on the homodyne phase setting.
    For $\sin \phi \approx 0$, SPACS are identified as nonclassical in a wide range of $\alpha$; cf. Fig. \ref{fig:performance_spacs}(b) for the detailed NN predictions for this case. 
    On the other hand, for suboptimal directions, SPACS are rarely recognized as nonclassical by the NN (except for small $\alpha$).
    Also, for large $\alpha$, SPACS are generally classified as classical in all directions.
    As a comparison, we show the NN prediction for experimental homodyne data generated by coherent states in Fig. \ref{fig:performance_spacs}(c) for the same parameters as used in Fig. \ref{fig:performance_spacs}(b).
    The network correctly recognizes coherent states as classical.
    
    The phase-dependent behavior of the NN output for the experimental SPACS can be explained by the fact that, for $\sin \phi \approx 0$, the quadrature distributions differ significantly from the one produced by a coherent state, while for other directions, the corresponding quadrature distributions resemble closely the ones of coherent states \cite{zavatta_2004,zavatta_2005,filippov_2013}.
    For small $\alpha<0.5$, SPACS resemble single-photon states and are thus recognized as nonclassical at all quadrature angles [see $p(x)$ for $\alpha=0.32$ in Fig. \ref{fig:performance_spacs}]. 
    On the other hand, for large $\alpha$, the quadrature distribution of the SPACS approaches the one of coherent states also in the optimal direction ($\phi=0$), and therefore, the NN eventually does not indicate nonclassicality anymore. In this regime, it is known that SPACS can be a good approximation of a coherent state of a larger amplitude \cite{jeong_2014}.
    The similarity of the SPACS quadrature distribution $p(x)$ for large $\alpha$ and the distribution from a coherent state explains the difficulty for the NN to classify SPACS as nonclassical in this regime.
    
    To summarize, for an optimal homodyne phase setting, SPACS are identified as nonclassical in a wide range of parameters. 
    It is a direct and simple method for testing nonclassicality of SPACS directly based on quadrature distributions.
    As we discuss below, this identification is successful even in a parameter regime where the homodyne distribution does not show sub-shot-noise or similarity to Fock states. 
    Therefore, the NN prediction proves operational for several different states and nonclassicality features.

\section{Influence of the training set and application to untrained data}
\label{sec:untrained}

In this section, we first discuss the ability of the NN to recognize different features of nonclassicality at the same time. Then, we test its performance to recognize nonclassicality of states that were not seen in the training phase and of measurement data consisting of varying sample sizes.

\subsection{Beyond single-feature recognition}
\begin{figure}[t]
		\center
		\includegraphics[width = 0.9\columnwidth]{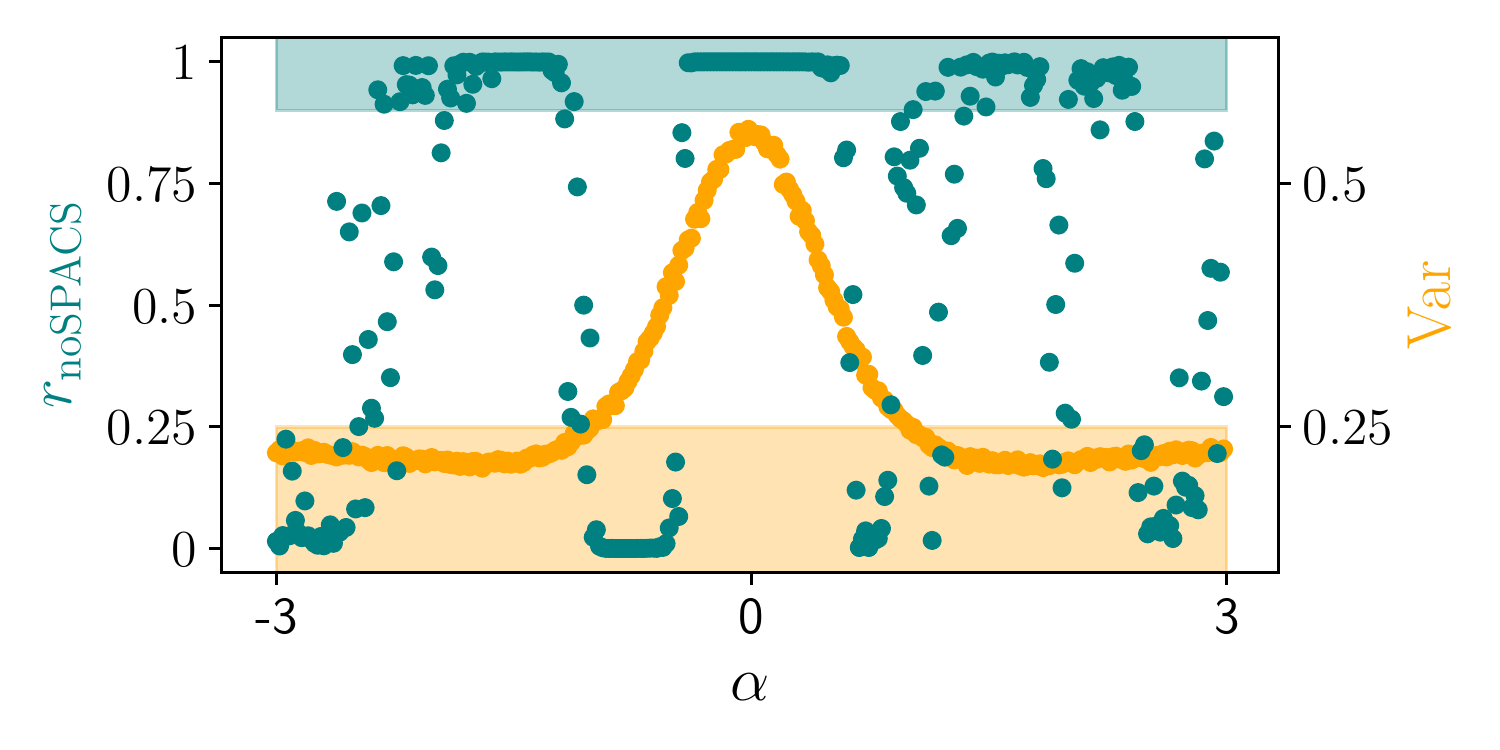}
		\caption{
		    Nonclassicality prediction of a network trained without single-photon-added coherent states (SPACS; $r_\mathrm{noSPACS}$, teal) for simulated SPACS as a function of $\alpha$ ($\phi=0$), together with the quadrature variance of the corresponding distributions (yellow).
		    Shaded regions are indicating nonclassicality seen by the network (teal) and by the variance criterion (yellow).
		}
		\label{fig:training_nospacs}
	\end{figure}

    To get some insights into which features are learned by the NN, we examine the performance in recognizing simulated SPACS of a network trained without SPACS; see Fig. \ref{fig:training_nospacs}. 
    We observe that a network which is not trained with SPACS recognizes the latter only in specific parameter regions (teal dots).
    For $|\alpha|\in[0,0.5]$, SPACS are recognized as nonclassical states due to their similarity to single-photon states. 
    On the other hand, in the parameter domain $|\alpha|\in[1,2]$, their nonclassicality is recognized because the variance of the quadrature distribution is significantly smaller than the vacuum variance.
    Beyond that, the distribution does not resemble Fock states and has a large quadrature variance and is, therefore, not classified as nonclassical. 
    For $|\alpha|>3$, the variance approaches the vacuum variance, making a correct classification as nonclassical impossible.
    
    In total, we see that the network can identify some SPACS even if they were not part of the training set.
    The network effectively identifies similarity to Fock states and sub-shotnoise variances.
    This is one example of the general fact that common features can lead to the identification of untrained data.
    In comparison, a NN that also used SPACS for its training can only achieve its performance (c.f. Fig. \ref{fig:performance_training}) by learning how to recognize similarity to SPACS where they do not resemble Fock or squeezed states.
    Therefore, we conclude that the network is sensitive to different nonclassical features at the same time and, thus, identifies nonclassicality beyond single features.
    Hence, a properly trained network can be advantageous, as it can recognize different nonclassical features for which one would otherwise need to implement different test conditions.
    This is particularly useful if the nonclassical features of the state to be tested are unknown. 
    As we have just seen, a state must not be part of the training set to be recognized by the network.
    The above analysis also indicates the necessity to train a deep NN to perform this task since simple baseline models like, e.g., sub-shot-noise variance only provide single-feature recognition.

\subsection{Application to untrained data}\label{sec:5}

    Now we discuss the performance of the NN on states which are not used in the training. 
    We apply the network to the family of so-called (odd) cat states
    \begin{equation}
    \left|\alpha_-\right\rangle=\frac{1}{\sqrt{2-2\exp(-2|\alpha|^2)}}\left(\left|\alpha\right\rangle-\left|-\alpha \right\rangle\right),
    \end{equation}
    where $\alpha\in\mathbb{R}^+$.
    As all states in this family consist of a coherent superposition of coherent states, they are all nonclassical.
    
\begin{figure}[t]
		\center
		\includegraphics[width = 0.9\columnwidth]{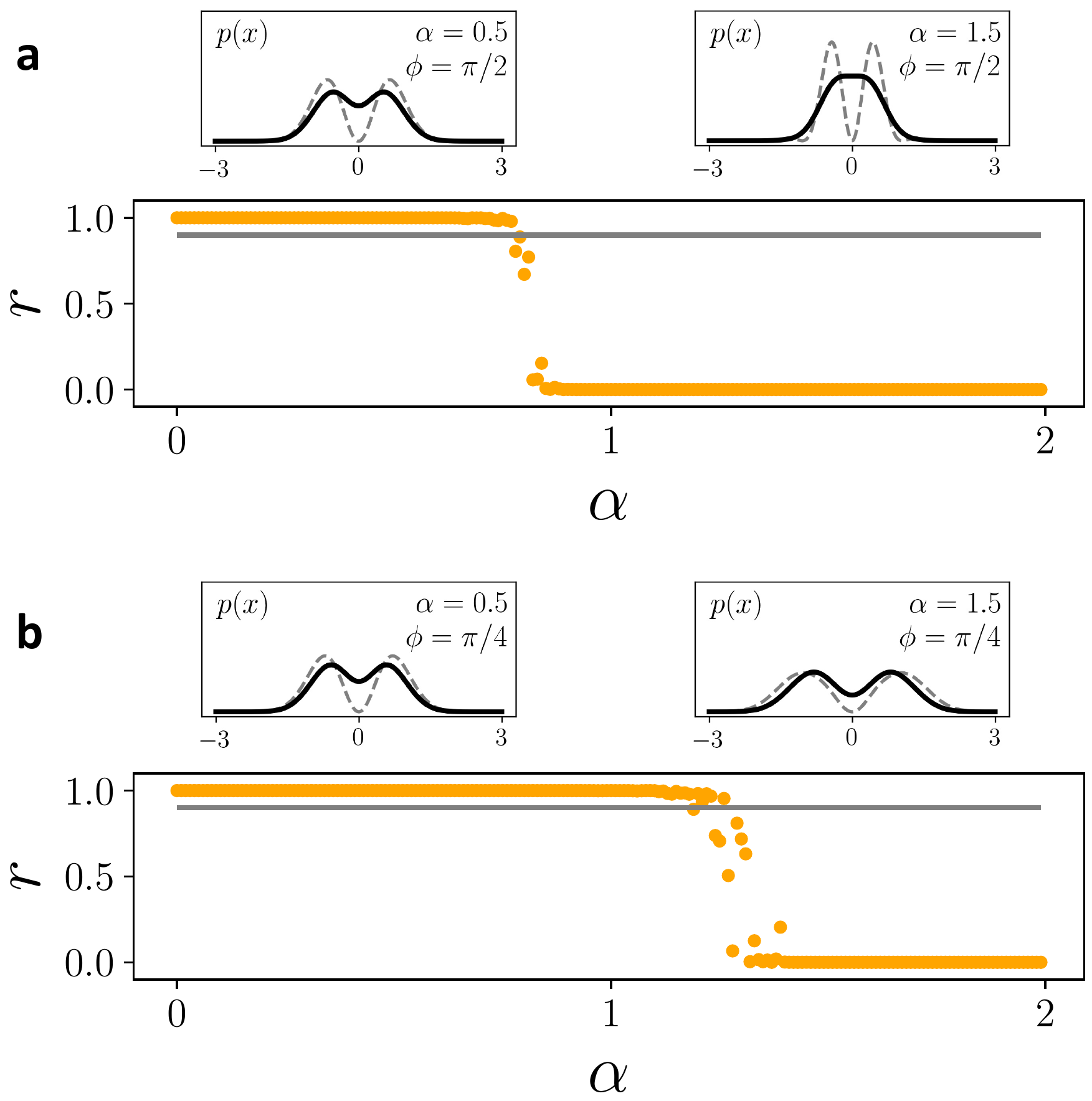}
		\caption{
        Prediction $r$ of the neural network (NN) for simulated quadrature measurements of the cat state $\left|\alpha_-\right\rangle$ as a function of $\alpha$, for (a) $\phi=\pi/2$ and for (b) $\phi=\pi/4$. Above the subplots, we show the quadrature distribution $p(x)$ for different $\alpha$ (solid), in comparison with the quadrature distribution using the same parameters but a quantum efficiency $\eta=1$ (dashed).}	\label{fig:performance_cat1}
    \end{figure}

    In Fig. \ref{fig:performance_cat1}, we show the $\alpha$-dependent prediction $r$ of the NN for quadrature measurements simulated for $\left|\alpha_-\right\rangle$. We use quadrature angles (a) $\phi=\pi/2$ and (b) $\phi=\pi/4$. 
    For each subfigure, we additionally display the quadrature distribution $p(x)$ for different values of $\alpha$ (solid) compared with $p(x)$ for the same parameters but using a quantum efficiency $\eta=1$ (dashed). 
    For both quadrature angles, the network correctly classifies the states as nonclassical in a significant range of $\alpha$.
    Thus, this example shows that the NN can certify nonclassicality also of untrained states.
    For larger $\alpha$, cat states are not identified as nonclassical.
    This behavior can be explained as follows. 
    For small $\alpha$, the cat state resembles a single-photon Fock state and can therefore be identified as nonclassical.
    For larger $\alpha$ and measured along $\phi=\pi/2$ with unit quantum efficiency $\eta=1$, the quadrature distribution develops a nonclassical interference pattern (a, dashed).
    However, for a realistic efficiency $\eta=0.6$, this interference is smoothed away (a, solid) such that the states are eventually classified as classical.
    Surprisingly, by choosing a different quadrature angle of, e.g.,  $\phi=\pi/4$, the cat states are classified as nonclassical in a wider range of $\alpha$ [Fig. \ref{fig:performance_cat1}(b)]. 
    This is because the quadrature distribution still resembles a Fock state in this direction. 
    Note that the performance of the NN prediction for cat states can be increased by including this family in the training process. 
    
    In summary, the NN is able to identify the nonclassicality also for states that were not used in the training process.
    However, for an optimized performance, it remains practical to adapt classes of states and parameter ranges in the training, see Appendix \ref{sec:appendix_traingdata}.

\subsection{Influence of the sample size}

\begin{figure}[t]
		\center
		\includegraphics[width = 0.9\columnwidth]{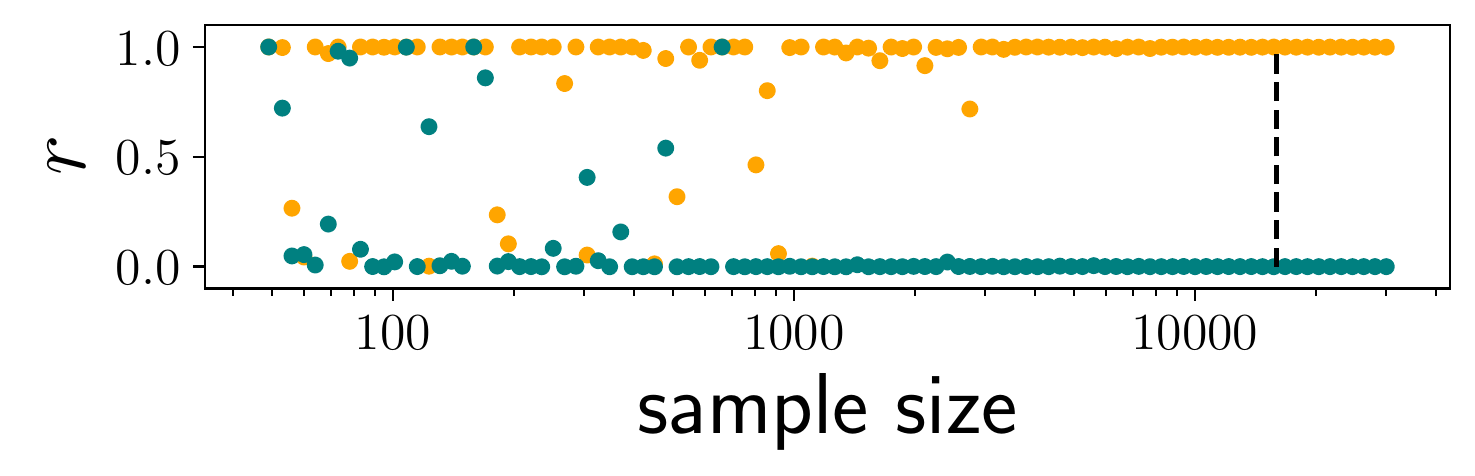}
		\caption{
        Prediction $r$ of the neural network (NN) for experimental data from single-photon-added coherent states (SPACS; yellow) and coherent states (teal) for $\alpha=0.32$ and $\phi=0$ as a function of the measurement sample size. The dashed vertical line indicates the sample size $16000$ that was used in the training phase.
        }
		\label{fig:samplesize}
    \end{figure}

Finally, we want to discuss the prediction of the NN if it is given measurement data with a smaller sample size than that used in the training phase. 
In Fig. \ref{fig:samplesize}, we show the NN nonclassicality prediction $r$ for experimental quadrature data of a SPACS (yellow) and a coherent state (teal) for $\alpha=0.32$ and $\phi=0$. 
We observe that a NN trained with sample sizes of $16000$ (dashed line) can correctly classify these two states for measurement data starting from sample sizes as low as $\sim 800$.
Decreasing the sample size even further results in false classification of coherent states as nonclassical and vice versa, which renders the NN prediction unreliable in this regime.

This analysis shows the flexibility of the NN even once it has been trained.
Importantly, the NN can provide conclusive predictions based on comparably very small sample sizes, which opens the possibility of online classification during measurements or fast (pre-)classification of data.
Note that the performance of the NN for small sample sizes can also be improved by training it with the corresponding sample size.

\section{Conclusions}\label{sec:6}
In this paper, we introduced an artificial NN-based nonclassicality identifier for single-mode quantum states of light measured with balanced homodyne detection. 
We trained the network using simulated homodyne detection data for realistic noisy measurements of different classical and nonclassical states. 
We observed that the trained network can correctly classify different classical and nonclassical states, i.e., coherent states, squeezed states, and SPACS, from real experimental data.
Furthermore, the network recognizes certain nonclassical states that were not used in the training phase of the network.
Compared with typical nonclassicality conditions based on homodyne tomography or other more complex nonclassicality tests, the strength of our approach lies in its simple implementation and the fact that only a small amount of data is required.
We would like to emphasize that the NN nonclassicality prediction cannot certify nonclassicality and, if necessary, should be complemented by an error-proof nonclassicality witness.

The ML-based classification offers a fast and accessible method to sort and preselect experimental data, considering that it circumvents the need to first perform homodyne tomography or the calculation of complex test conditions and, as we showed, performs well also on small sample sizes. It is furthermore easy to implement and applicable in multiple experimental settings.
ML has been used before for the detection of quantum effects \cite{gebhart_2020,cimini_2020,tiunov_2020}. In this context, it is important to highlight that the presented approach can detect phase-sensitive nonclassical features, which was not possible with previous results \cite{gebhart_2020}.

Further, the network approach can be used to search interesting experimental parameter regimes, especially if the production rate of detection events is small.
To maximize the accuracy of the NN prediction in experiments, any specific information about possible states and noise (such as phase or amplitude noise) should be included in the training phase.
Finally, note that the presented approach can be generalized to multi-mode scenarios and might be adapted to the identification of entanglement in a similar fashion.
Also, different additional ML methods such as convolutional layers or regularizations can be considered to optimize the performance of the NN nonclassicality prediction and make it more applicable to untrained data.

\section*{Acknowledgements}

    The authors thank B. Hage for kindly providing the experimental quadrature data for the squeezed states.
    E.A. acknowledges funding from the European Union's (EU's) Horizon 2020 Research and Innovation Programme under the Marie Sk\l{}odowska-Curie IF (InDiQE - EU project No. 845486). N.B., M.B., and A.Z. acknowledge funding from the EU under the ERA-NET QuantERA project "ShoQC" and the FET Flagship on Quantum Technologies project "Qombs" (grant no. 820419).

\appendix
\section*{Appendix}

\section{Choice of classical states in the training phase}\label{sec:appendix_traingdata}

Here, we want to emphasize the role of the choice of different classical states in the training data. In the main text, we used, in addition to coherent and thermal states, mixtures of coherent states of the form $\rho_\mathrm{mix}=(\ket{\alpha}\bra{\alpha}+\ket{-\alpha}\bra{-\alpha})/2$, where $\alpha$ is sampled from the same parameter domain as for coherent states. 
These states have to be included in the training because, otherwise, training only on classical states with single-peaked quadrature distributions, the NN might interpret double- or multi-peak structures as features of nonclassicality. 
However, the choice of which classical state to use here is not unique. 
For instance, a different classical state that occurs typically in experiments is a phase averaged coherent state, $\rho_\mathrm{av}=\int_0^{2\pi}\mathrm{d}\phi\ket{\alpha e^{i\phi}}\bra{\alpha e^{i\phi}}/2\pi$.
Using $\rho_\mathrm{av}$ instead of $\rho_\mathrm{mix}$ in the training results in a similar performance of the NN as in the main text, with the expectation that, for larger $\alpha$, $\rho_\mathrm{mix}$ (and therefore also cat states measured along $\phi=0$) are classified as nonclassical. 

This points to an important caveat of the NN classification of nonclassicality: 
as mentioned in the main text, the different states used in the training phase must be chosen carefully, given the experimental conditions. 
Training with more families of classical states decreases the probability of false identification of nonclassicality for states that were not seen in the training. 
At the same time, it makes it harder for the NN to learn nonclassicality features of the corresponding nonclassical training states. 
This discussion shows that the NN nonclassicality classification, while representing a simple and fast nonclassicality identification if possible input states are known, is not universal and does not yield a strict nonclassicality certification. 

\section{Parameters and probabilities used for the simulation of quadrature measurement data}\label{sec:Appendix}

    Here, we specify the state-dependent quadrature probability distributions and the corresponding parameters used in the simulation of quadrature measurement data for the different states in the main text.
    In Table \ref{tab:statestatistics}, we list the different states together with the corresponding parameter regions used in the simulations and the quadrature distribution along the quadrature angle $\phi$. Note that we use a vacuum variance of $1/4$. 
    
    For the simulation of the training data, we fixed a quadrature angle $\phi=0$.
    For thermal and Fock states, this restriction does not influence the distribution, as these states are phase insensitive.
    For coherent states with amplitude $\alpha$, the distribution along a nonzero $\phi$ is equivalent to the one of a coherent state with amplitude $\alpha\cos\phi$, measured along a zero quadrature angle.
    For squeezed coherent states and SPACS, this choice assures that only quadrature distributions which show nonclassical features are used in the training.

    As noted in Ref.~\cite{Note1}, the different parameter limits are chosen such that the probability for an event outside the considered measurement range, $\left|x\right|>8$, is small ($<10^{-6}$).
    Note that, for SPACS, we further restrict the parameters ($|\alpha|\leq 3$) to a domain where the network is able to separate them clearly from the classical states. For the simulation of the squeezed states, the squeezing parameter is chosen uniformly in $\xi\in[0.5,1]$.

\begin{table*}[b]
    \centering
    \begin{tabular}{ccc}
    \hline \hline 
      State & Parameters & Probability $p(x,\phi)$   \\ \hline 
      coherent   & $\alpha\in [-5,5]$ & $\sqrt{\frac{2}{\pi}}\exp[-2(x-\sqrt{\eta}\alpha\cos\phi)^2]$ \\   
      thermal   & $\bar n \in [0,5]$  & $\sqrt{\frac{2}{\pi(1+2\eta\bar n)}}\exp\left[-\frac{2x^2}{1+2\eta\bar n}\right]$  \\  
      Fock   & $n\in\{1,\dots,6\}$ & $\sqrt{\frac{2}{\pi}}\sum_{k=0}^n \binom{n}{k}\frac{\eta^k}{2^k k!}e^{-2 x^2}H_{2k}(\sqrt{2}x)$ \\  
      squeezed coherent  & $\alpha\in [-5,5],\xi\in [0.5,1]$ & $\sqrt{\frac{2}{\pi(1-\eta+e^{2|\xi|}\cos^2\phi+e^{-2|\xi|}\sin^2\phi)}}\exp \left[-\frac{2(x-\sqrt{\eta}\alpha)^2}{1-\eta+e^{2|\xi|}\cos^2\phi+e^{-2|\xi|}\sin^2\phi}\right]$ \\ 
      SPACS   & $\alpha\in [-3,3]$ & $\frac{1}{1+\alpha^2}\sqrt{\frac{2}{\pi}}\exp[-2(x-\sqrt{\eta}\alpha\cos\phi)^2]$  \\  
       & & $\times \left[\eta\left(2x\cos\phi-\frac{2\eta-1}{\sqrt{\eta}}\alpha\right)^2+4\eta x^2\sin^2 \phi +(1-\eta)(1+4\eta\alpha^2\sin^2\phi)\right]$ \cite{zavatta_2005}\\
      cat & $\alpha\in [-5,5]$ & $ \sqrt{\frac{2}{\pi}}\frac{1}{2-2e^{-2\alpha^2}}\Big\{ \exp[-2(x-\sqrt{\eta}\alpha\cos\phi)^2]$ \\
      & & $- 2e^{-2\alpha^2}\operatorname{Re}[\exp[-2(x+i\sqrt{\eta}\alpha\sin\phi)^2]]+  \exp[-2(x+\sqrt{\eta}\alpha\cos\phi)^2]\Big\}$ \\ 
      \hline \hline 
    \end{tabular}
    \caption{
    For each family of states, the used parameter regions and the corresponding quadrature distributions $p(x,\phi)$ are shown, where $x$ is the quadrature value and $\phi$ is the phase in the balanced homodyne detection. $\eta$ is the overall quantum efficiency. 
    }
    \label{tab:statestatistics}
\end{table*}

%


\begin{thebibliography}{64}%
\makeatletter
\providecommand \@ifxundefined [1]{%
 \@ifx{#1\undefined}
}%
\providecommand \@ifnum [1]{%
 \ifnum #1\expandafter \@firstoftwo
 \else \expandafter \@secondoftwo
 \fi
}%
\providecommand \@ifx [1]{%
 \ifx #1\expandafter \@firstoftwo
 \else \expandafter \@secondoftwo
 \fi
}%
\providecommand \natexlab [1]{#1}%
\providecommand \enquote  [1]{``#1''}%
\providecommand \bibnamefont  [1]{#1}%
\providecommand \bibfnamefont [1]{#1}%
\providecommand \citenamefont [1]{#1}%
\providecommand \href@noop [0]{\@secondoftwo}%
\providecommand \href [0]{\begingroup \@sanitize@url \@href}%
\providecommand \@href[1]{\@@startlink{#1}\@@href}%
\providecommand \@@href[1]{\endgroup#1\@@endlink}%
\providecommand \@sanitize@url [0]{\catcode `\\12\catcode `\$12\catcode
  `\&12\catcode `\#12\catcode `\^12\catcode `\_12\catcode `\%12\relax}%
\providecommand \@@startlink[1]{}%
\providecommand \@@endlink[0]{}%
\providecommand \url  [0]{\begingroup\@sanitize@url \@url }%
\providecommand \@url [1]{\endgroup\@href {#1}{\urlprefix }}%
\providecommand \urlprefix  [0]{URL }%
\providecommand \Eprint [0]{\href }%
\providecommand \doibase [0]{http://dx.doi.org/}%
\providecommand \selectlanguage [0]{\@gobble}%
\providecommand \bibinfo  [0]{\@secondoftwo}%
\providecommand \bibfield  [0]{\@secondoftwo}%
\providecommand \translation [1]{[#1]}%
\providecommand \BibitemOpen [0]{}%
\providecommand \bibitemStop [0]{}%
\providecommand \bibitemNoStop [0]{.\EOS\space}%
\providecommand \EOS [0]{\spacefactor3000\relax}%
\providecommand \BibitemShut  [1]{\csname bibitem#1\endcsname}%
\let\auto@bib@innerbib\@empty
\bibitem [{\citenamefont {Braunstein}\ and\ \citenamefont {van
  Loock}(2005)}]{braunstein2005}%
  \BibitemOpen
  \bibfield  {author} {\bibinfo {author} {\bibfnamefont {Samuel~L.}\
  \bibnamefont {Braunstein}}\ and\ \bibinfo {author} {\bibfnamefont {Peter}\
  \bibnamefont {van Loock}},\ }\bibfield  {title} {\enquote {\bibinfo {title}
  {Quantum information with continuous variables},}\ }\href {\doibase
  10.1103/RevModPhys.77.513} {\bibfield  {journal} {\bibinfo  {journal} {Rev.
  Mod. Phys.}\ }\textbf {\bibinfo {volume} {77}},\ \bibinfo {pages} {513--577}
  (\bibinfo {year} {2005})}\BibitemShut {NoStop}%
\bibitem [{\citenamefont {Streltsov}\ \emph {et~al.}(2017)\citenamefont
  {Streltsov}, \citenamefont {Adesso},\ and\ \citenamefont
  {Plenio}}]{streltsov_2017}%
  \BibitemOpen
  \bibfield  {author} {\bibinfo {author} {\bibfnamefont {Alexander}\
  \bibnamefont {Streltsov}}, \bibinfo {author} {\bibfnamefont {Gerardo}\
  \bibnamefont {Adesso}}, \ and\ \bibinfo {author} {\bibfnamefont {Martin~B.}\
  \bibnamefont {Plenio}},\ }\bibfield  {title} {\enquote {\bibinfo {title}
  {Colloquium: Quantum coherence as a resource},}\ }\href {\doibase
  10.1103/RevModPhys.89.041003} {\bibfield  {journal} {\bibinfo  {journal}
  {Rev. Mod. Phys.}\ }\textbf {\bibinfo {volume} {89}},\ \bibinfo {pages}
  {041003} (\bibinfo {year} {2017})}\BibitemShut {NoStop}%
\bibitem [{\citenamefont {Sperling}\ and\ \citenamefont
  {Walmsley}(2018)}]{sperling_2018}%
  \BibitemOpen
  \bibfield  {author} {\bibinfo {author} {\bibfnamefont {J.}~\bibnamefont
  {Sperling}}\ and\ \bibinfo {author} {\bibfnamefont {I.~A.}\ \bibnamefont
  {Walmsley}},\ }\bibfield  {title} {\enquote {\bibinfo {title}
  {Quasiprobability representation of quantum coherence},}\ }\href {\doibase
  10.1103/PhysRevA.97.062327} {\bibfield  {journal} {\bibinfo  {journal} {Phys.
  Rev. A}\ }\textbf {\bibinfo {volume} {97}},\ \bibinfo {pages} {062327}
  (\bibinfo {year} {2018})}\BibitemShut {NoStop}%
\bibitem [{\citenamefont {Titlaer}\ and\ \citenamefont
  {Glauber}(1965)}]{titulaer_1965}%
  \BibitemOpen
  \bibfield  {author} {\bibinfo {author} {\bibfnamefont {U.~M.}\ \bibnamefont
  {Titlaer}}\ and\ \bibinfo {author} {\bibfnamefont {R.~J.}\ \bibnamefont
  {Glauber}},\ }\bibfield  {title} {\enquote {\bibinfo {title} {Correlation
  functions for coherent fields},}\ }\href {\doibase 10.1103/PhysRev.140.B676}
  {\bibfield  {journal} {\bibinfo  {journal} {Phys. Rev.}\ }\textbf {\bibinfo
  {volume} {140}},\ \bibinfo {pages} {B676--B682} (\bibinfo {year}
  {1965})}\BibitemShut {NoStop}%
\bibitem [{\citenamefont {Mandel}(1986)}]{mandel_1986}%
  \BibitemOpen
  \bibfield  {author} {\bibinfo {author} {\bibfnamefont {L}~\bibnamefont
  {Mandel}},\ }\bibfield  {title} {\enquote {\bibinfo {title} {Non-classical
  states of the electromagnetic field},}\ }\href {\doibase
  10.1088/0031-8949/1986/t12/005} {\bibfield  {journal} {\bibinfo  {journal}
  {Phys. Scr.}\ }\textbf {\bibinfo {volume} {T12}},\ \bibinfo {pages} {34--42}
  (\bibinfo {year} {1986})}\BibitemShut {NoStop}%
\bibitem [{\citenamefont {Glauber}(1963)}]{glauber_1963}%
  \BibitemOpen
  \bibfield  {author} {\bibinfo {author} {\bibfnamefont {Roy~J.}\ \bibnamefont
  {Glauber}},\ }\bibfield  {title} {\enquote {\bibinfo {title} {Coherent and
  incoherent states of the radiation field},}\ }\href {\doibase
  10.1103/PhysRev.131.2766} {\bibfield  {journal} {\bibinfo  {journal} {Phys.
  Rev.}\ }\textbf {\bibinfo {volume} {131}},\ \bibinfo {pages} {2766--2788}
  (\bibinfo {year} {1963})}\BibitemShut {NoStop}%
\bibitem [{\citenamefont {Sudarshan}(1963)}]{sudarshan_1963}%
  \BibitemOpen
  \bibfield  {author} {\bibinfo {author} {\bibfnamefont {E.~C.~G.}\
  \bibnamefont {Sudarshan}},\ }\bibfield  {title} {\enquote {\bibinfo {title}
  {Equivalence of semiclassical and quantum mechanical descriptions of
  statistical light beams},}\ }\href {\doibase 10.1103/PhysRevLett.10.277}
  {\bibfield  {journal} {\bibinfo  {journal} {Phys. Rev. Lett.}\ }\textbf
  {\bibinfo {volume} {10}},\ \bibinfo {pages} {277--279} (\bibinfo {year}
  {1963})}\BibitemShut {NoStop}%
\bibitem [{\citenamefont {Carmichael}\ and\ \citenamefont
  {Walls}(1976)}]{Carmichael_1976}%
  \BibitemOpen
  \bibfield  {author} {\bibinfo {author} {\bibfnamefont {H~J}\ \bibnamefont
  {Carmichael}}\ and\ \bibinfo {author} {\bibfnamefont {D~F}\ \bibnamefont
  {Walls}},\ }\bibfield  {title} {\enquote {\bibinfo {title} {Proposal for the
  measurement of the resonant stark effect by photon correlation techniques},}\
  }\href {\doibase 10.1088/0022-3700/9/4/001} {\bibfield  {journal} {\bibinfo
  {journal} {J. Phys. B}\ }\textbf {\bibinfo {volume} {9}},\ \bibinfo {pages}
  {L43--L46} (\bibinfo {year} {1976})}\BibitemShut {NoStop}%
\bibitem [{\citenamefont {Kimble}\ and\ \citenamefont
  {Mandel}(1976)}]{Kimble_1976}%
  \BibitemOpen
  \bibfield  {author} {\bibinfo {author} {\bibfnamefont {H.~J.}\ \bibnamefont
  {Kimble}}\ and\ \bibinfo {author} {\bibfnamefont {L.}~\bibnamefont
  {Mandel}},\ }\bibfield  {title} {\enquote {\bibinfo {title} {Theory of
  resonance fluorescence},}\ }\href {\doibase 10.1103/PhysRevA.13.2123}
  {\bibfield  {journal} {\bibinfo  {journal} {Phys. Rev. A}\ }\textbf {\bibinfo
  {volume} {13}},\ \bibinfo {pages} {2123--2144} (\bibinfo {year}
  {1976})}\BibitemShut {NoStop}%
\bibitem [{\citenamefont {Kimble}\ \emph {et~al.}(1977)\citenamefont {Kimble},
  \citenamefont {Dagenais},\ and\ \citenamefont {Mandel}}]{kimble_1977}%
  \BibitemOpen
  \bibfield  {author} {\bibinfo {author} {\bibfnamefont {H.~J.}\ \bibnamefont
  {Kimble}}, \bibinfo {author} {\bibfnamefont {M.}~\bibnamefont {Dagenais}}, \
  and\ \bibinfo {author} {\bibfnamefont {L.}~\bibnamefont {Mandel}},\
  }\bibfield  {title} {\enquote {\bibinfo {title} {Photon antibunching in
  resonance fluorescence},}\ }\href {\doibase 10.1103/PhysRevLett.39.691}
  {\bibfield  {journal} {\bibinfo  {journal} {Phys. Rev. Lett.}\ }\textbf
  {\bibinfo {volume} {39}},\ \bibinfo {pages} {691--695} (\bibinfo {year}
  {1977})}\BibitemShut {NoStop}%
\bibitem [{\citenamefont {Mandel}(1979)}]{mandel_1979}%
  \BibitemOpen
  \bibfield  {author} {\bibinfo {author} {\bibfnamefont {L.}~\bibnamefont
  {Mandel}},\ }\bibfield  {title} {\enquote {\bibinfo {title} {Sub-poissonian
  photon statistics in resonance fluorescence},}\ }\href {\doibase
  10.1364/OL.4.000205} {\bibfield  {journal} {\bibinfo  {journal} {Opt. Lett.}\
  }\textbf {\bibinfo {volume} {4}},\ \bibinfo {pages} {205--207} (\bibinfo
  {year} {1979})}\BibitemShut {NoStop}%
\bibitem [{\citenamefont {Zou}\ and\ \citenamefont {Mandel}(1990)}]{Zou_1990}%
  \BibitemOpen
  \bibfield  {author} {\bibinfo {author} {\bibfnamefont {X.~T.}\ \bibnamefont
  {Zou}}\ and\ \bibinfo {author} {\bibfnamefont {L.}~\bibnamefont {Mandel}},\
  }\bibfield  {title} {\enquote {\bibinfo {title} {Photon-antibunching and
  sub-poissonian photon statistics},}\ }\href {\doibase
  10.1103/PhysRevA.41.475} {\bibfield  {journal} {\bibinfo  {journal} {Phys.
  Rev. A}\ }\textbf {\bibinfo {volume} {41}},\ \bibinfo {pages} {475--476}
  (\bibinfo {year} {1990})}\BibitemShut {NoStop}%
\bibitem [{\citenamefont {Yuen}(1976)}]{Yuen_1976}%
  \BibitemOpen
  \bibfield  {author} {\bibinfo {author} {\bibfnamefont {Horace~P.}\
  \bibnamefont {Yuen}},\ }\bibfield  {title} {\enquote {\bibinfo {title}
  {Two-photon coherent states of the radiation field},}\ }\href {\doibase
  10.1103/PhysRevA.13.2226} {\bibfield  {journal} {\bibinfo  {journal} {Phys.
  Rev. A}\ }\textbf {\bibinfo {volume} {13}},\ \bibinfo {pages} {2226--2243}
  (\bibinfo {year} {1976})}\BibitemShut {NoStop}%
\bibitem [{\citenamefont {Walls}(1983)}]{Walls_1983}%
  \BibitemOpen
  \bibfield  {author} {\bibinfo {author} {\bibfnamefont {D.~F.}\ \bibnamefont
  {Walls}},\ }\bibfield  {title} {\enquote {\bibinfo {title} {Squeezed states
  of light},}\ }\href {\doibase 10.1038/306141a0} {\bibfield  {journal}
  {\bibinfo  {journal} {Nature}\ }\textbf {\bibinfo {volume} {306}},\ \bibinfo
  {pages} {141--146} (\bibinfo {year} {1983})}\BibitemShut {NoStop}%
\bibitem [{\citenamefont {Caves}\ and\ \citenamefont
  {Schumaker}(1985)}]{Caves_1985}%
  \BibitemOpen
  \bibfield  {author} {\bibinfo {author} {\bibfnamefont {Carlton~M.}\
  \bibnamefont {Caves}}\ and\ \bibinfo {author} {\bibfnamefont {Bonny~L.}\
  \bibnamefont {Schumaker}},\ }\bibfield  {title} {\enquote {\bibinfo {title}
  {New formalism for two-photon quantum optics. i. quadrature phases and
  squeezed states},}\ }\href {\doibase 10.1103/PhysRevA.31.3068} {\bibfield
  {journal} {\bibinfo  {journal} {Phys. Rev. A}\ }\textbf {\bibinfo {volume}
  {31}},\ \bibinfo {pages} {3068--3092} (\bibinfo {year} {1985})}\BibitemShut
  {NoStop}%
\bibitem [{\citenamefont {Loudon}\ and\ \citenamefont
  {Knight}(1987)}]{Loudon_1987}%
  \BibitemOpen
  \bibfield  {author} {\bibinfo {author} {\bibfnamefont {R.}~\bibnamefont
  {Loudon}}\ and\ \bibinfo {author} {\bibfnamefont {P.L.}\ \bibnamefont
  {Knight}},\ }\bibfield  {title} {\enquote {\bibinfo {title} {Squeezed
  light},}\ }\href {\doibase 10.1080/09500348714550721} {\bibfield  {journal}
  {\bibinfo  {journal} {J. Mod. Opt.}\ }\textbf {\bibinfo {volume} {34}},\
  \bibinfo {pages} {709--759} (\bibinfo {year} {1987})}\BibitemShut {NoStop}%
\bibitem [{\citenamefont {Dodonov}(2002)}]{Dodonov_2002}%
  \BibitemOpen
  \bibfield  {author} {\bibinfo {author} {\bibfnamefont {V~V}\ \bibnamefont
  {Dodonov}},\ }\bibfield  {title} {\enquote {\bibinfo {title} {`nonclassical'
  states in quantum optics: a `squeezed' review of the first 75 years},}\
  }\href {\doibase 10.1088/1464-4266/4/1/201} {\bibfield  {journal} {\bibinfo
  {journal} {J. Opt. B: Quantum Semiclass. Opt.}\ }\textbf {\bibinfo {volume}
  {4}},\ \bibinfo {pages} {R1--R33} (\bibinfo {year} {2002})}\BibitemShut
  {NoStop}%
\bibitem [{\citenamefont {Vogel}\ and\ \citenamefont
  {Sperling}(2014)}]{vogel_2014}%
  \BibitemOpen
  \bibfield  {author} {\bibinfo {author} {\bibfnamefont {W.}~\bibnamefont
  {Vogel}}\ and\ \bibinfo {author} {\bibfnamefont {J.}~\bibnamefont
  {Sperling}},\ }\bibfield  {title} {\enquote {\bibinfo {title} {Unified
  quantification of nonclassicality and entanglement},}\ }\href {\doibase
  10.1103/PhysRevA.89.052302} {\bibfield  {journal} {\bibinfo  {journal} {Phys.
  Rev. A}\ }\textbf {\bibinfo {volume} {89}},\ \bibinfo {pages} {052302}
  (\bibinfo {year} {2014})}\BibitemShut {NoStop}%
\bibitem [{\citenamefont {Killoran}\ \emph {et~al.}(2016)\citenamefont
  {Killoran}, \citenamefont {Steinhoff},\ and\ \citenamefont
  {Plenio}}]{killoran_2016}%
  \BibitemOpen
  \bibfield  {author} {\bibinfo {author} {\bibfnamefont {N.}~\bibnamefont
  {Killoran}}, \bibinfo {author} {\bibfnamefont {F.~E.~S.}\ \bibnamefont
  {Steinhoff}}, \ and\ \bibinfo {author} {\bibfnamefont {M.~B.}\ \bibnamefont
  {Plenio}},\ }\bibfield  {title} {\enquote {\bibinfo {title} {Converting
  nonclassicality into entanglement},}\ }\href {\doibase
  10.1103/PhysRevLett.116.080402} {\bibfield  {journal} {\bibinfo  {journal}
  {Phys. Rev. Lett.}\ }\textbf {\bibinfo {volume} {116}},\ \bibinfo {pages}
  {080402} (\bibinfo {year} {2016})}\BibitemShut {NoStop}%
\bibitem [{\citenamefont {Welsch}\ \emph {et~al.}(1999)\citenamefont {Welsch},
  \citenamefont {Vogel},\ and\ \citenamefont {Opatrn\'y}}]{welsch_1999}%
  \BibitemOpen
  \bibfield  {author} {\bibinfo {author} {\bibfnamefont {Dirk-Gunnar}\
  \bibnamefont {Welsch}}, \bibinfo {author} {\bibfnamefont {Werner}\
  \bibnamefont {Vogel}}, \ and\ \bibinfo {author} {\bibfnamefont {Tom\'a\v{s}}\
  \bibnamefont {Opatrn\'y}},\ }\href {\doibase
  https://doi.org/10.1016/S0079-6638(08)70389-5} {\emph {\bibinfo {title}
  {Homodyne Detection and Quantum-State Reconstruction}}},\ edited by\ \bibinfo
  {editor} {\bibfnamefont {E.}~\bibnamefont {Wolf}},\ \bibinfo {series}
  {Progress in Optics}, Vol.~\bibinfo {volume} {39}\ (\bibinfo  {publisher}
  {Elsevier},\ \bibinfo {year} {1999})\ pp.\ \bibinfo {pages} {63 --
  211}\BibitemShut {NoStop}%
\bibitem [{\citenamefont {Paris}\ and\ \citenamefont
  {\v{R}eh\'{a}\v{c}ek}(2004)}]{paris_2004}%
  \BibitemOpen
  \bibfield  {author} {\bibinfo {author} {\bibfnamefont {M.}~\bibnamefont
  {Paris}}\ and\ \bibinfo {author} {\bibfnamefont {J.}~\bibnamefont
  {\v{R}eh\'{a}\v{c}ek}},\ }\href@noop {} {\emph {\bibinfo {title} {Quantum
  State Estimation}}}\ (\bibinfo  {publisher} {Springer},\ \bibinfo {year}
  {2004})\BibitemShut {NoStop}%
\bibitem [{\citenamefont {Lvovsky}\ and\ \citenamefont
  {Raymer}(2009)}]{lvovsky_2009}%
  \BibitemOpen
  \bibfield  {author} {\bibinfo {author} {\bibfnamefont {A.~I.}\ \bibnamefont
  {Lvovsky}}\ and\ \bibinfo {author} {\bibfnamefont {M.~G.}\ \bibnamefont
  {Raymer}},\ }\bibfield  {title} {\enquote {\bibinfo {title}
  {Continuous-variable optical quantum-state tomography},}\ }\href {\doibase
  10.1103/RevModPhys.81.299} {\bibfield  {journal} {\bibinfo  {journal} {Rev.
  Mod. Phys.}\ }\textbf {\bibinfo {volume} {81}},\ \bibinfo {pages} {299--332}
  (\bibinfo {year} {2009})}\BibitemShut {NoStop}%
\bibitem [{\citenamefont {Cahill}\ and\ \citenamefont
  {Glauber}(1969)}]{cahill_1969b}%
  \BibitemOpen
  \bibfield  {author} {\bibinfo {author} {\bibfnamefont {K.~E.}\ \bibnamefont
  {Cahill}}\ and\ \bibinfo {author} {\bibfnamefont {R.~J.}\ \bibnamefont
  {Glauber}},\ }\bibfield  {title} {\enquote {\bibinfo {title} {Density
  operators and quasiprobability distributions},}\ }\href {\doibase
  10.1103/PhysRev.177.1882} {\bibfield  {journal} {\bibinfo  {journal} {Phys.
  Rev.}\ }\textbf {\bibinfo {volume} {177}},\ \bibinfo {pages} {1882--1902}
  (\bibinfo {year} {1969})}\BibitemShut {NoStop}%
\bibitem [{\citenamefont {Kiesel}\ \emph {et~al.}(2008)\citenamefont {Kiesel},
  \citenamefont {Vogel}, \citenamefont {Parigi}, \citenamefont {Zavatta},\ and\
  \citenamefont {Bellini}}]{Kiesel_2008}%
  \BibitemOpen
  \bibfield  {author} {\bibinfo {author} {\bibfnamefont {T.}~\bibnamefont
  {Kiesel}}, \bibinfo {author} {\bibfnamefont {W.}~\bibnamefont {Vogel}},
  \bibinfo {author} {\bibfnamefont {V.}~\bibnamefont {Parigi}}, \bibinfo
  {author} {\bibfnamefont {A.}~\bibnamefont {Zavatta}}, \ and\ \bibinfo
  {author} {\bibfnamefont {M.}~\bibnamefont {Bellini}},\ }\bibfield  {title}
  {\enquote {\bibinfo {title} {Experimental determination of a nonclassical
  glauber-sudarshan $p$ function},}\ }\href {\doibase
  10.1103/PhysRevA.78.021804} {\bibfield  {journal} {\bibinfo  {journal} {Phys.
  Rev. A}\ }\textbf {\bibinfo {volume} {78}},\ \bibinfo {pages} {021804(R)}
  (\bibinfo {year} {2008})}\BibitemShut {NoStop}%
\bibitem [{\citenamefont {Smithey}\ \emph {et~al.}(1993)\citenamefont
  {Smithey}, \citenamefont {Beck}, \citenamefont {Raymer},\ and\ \citenamefont
  {Faridani}}]{smithey_1993}%
  \BibitemOpen
  \bibfield  {author} {\bibinfo {author} {\bibfnamefont {D.~T.}\ \bibnamefont
  {Smithey}}, \bibinfo {author} {\bibfnamefont {M.}~\bibnamefont {Beck}},
  \bibinfo {author} {\bibfnamefont {M.~G.}\ \bibnamefont {Raymer}}, \ and\
  \bibinfo {author} {\bibfnamefont {A.}~\bibnamefont {Faridani}},\ }\bibfield
  {title} {\enquote {\bibinfo {title} {Measurement of the wigner distribution
  and the density matrix of a light mode using optical homodyne tomography:
  Application to squeezed states and the vacuum},}\ }\href {\doibase
  10.1103/PhysRevLett.70.1244} {\bibfield  {journal} {\bibinfo  {journal}
  {Phys. Rev. Lett.}\ }\textbf {\bibinfo {volume} {70}},\ \bibinfo {pages}
  {1244--1247} (\bibinfo {year} {1993})}\BibitemShut {NoStop}%
\bibitem [{\citenamefont {Dunn}\ \emph {et~al.}(1995)\citenamefont {Dunn},
  \citenamefont {Walmsley},\ and\ \citenamefont {Mukamel}}]{Dunn_1995}%
  \BibitemOpen
  \bibfield  {author} {\bibinfo {author} {\bibfnamefont {T.~J.}\ \bibnamefont
  {Dunn}}, \bibinfo {author} {\bibfnamefont {I.~A.}\ \bibnamefont {Walmsley}},
  \ and\ \bibinfo {author} {\bibfnamefont {S.}~\bibnamefont {Mukamel}},\
  }\bibfield  {title} {\enquote {\bibinfo {title} {Experimental determination
  of the quantum-mechanical state of a molecular vibrational mode using
  fluorescence tomography},}\ }\href {\doibase 10.1103/PhysRevLett.74.884}
  {\bibfield  {journal} {\bibinfo  {journal} {Phys. Rev. Lett.}\ }\textbf
  {\bibinfo {volume} {74}},\ \bibinfo {pages} {884--887} (\bibinfo {year}
  {1995})}\BibitemShut {NoStop}%
\bibitem [{\citenamefont {Leibfried}\ \emph {et~al.}(1996)\citenamefont
  {Leibfried}, \citenamefont {Meekhof}, \citenamefont {King}, \citenamefont
  {Monroe}, \citenamefont {Itano},\ and\ \citenamefont
  {Wineland}}]{leibfried_1996}%
  \BibitemOpen
  \bibfield  {author} {\bibinfo {author} {\bibfnamefont {D.}~\bibnamefont
  {Leibfried}}, \bibinfo {author} {\bibfnamefont {D.~M.}\ \bibnamefont
  {Meekhof}}, \bibinfo {author} {\bibfnamefont {B.~E.}\ \bibnamefont {King}},
  \bibinfo {author} {\bibfnamefont {C.}~\bibnamefont {Monroe}}, \bibinfo
  {author} {\bibfnamefont {W.~M.}\ \bibnamefont {Itano}}, \ and\ \bibinfo
  {author} {\bibfnamefont {D.~J.}\ \bibnamefont {Wineland}},\ }\bibfield
  {title} {\enquote {\bibinfo {title} {Experimental determination of the
  motional quantum state of a trapped atom},}\ }\href {\doibase
  10.1103/PhysRevLett.77.4281} {\bibfield  {journal} {\bibinfo  {journal}
  {Phys. Rev. Lett.}\ }\textbf {\bibinfo {volume} {77}},\ \bibinfo {pages}
  {4281--4285} (\bibinfo {year} {1996})}\BibitemShut {NoStop}%
\bibitem [{\citenamefont {Del{\'e}glise}\ \emph {et~al.}(2008)\citenamefont
  {Del{\'e}glise}, \citenamefont {Dotsenko}, \citenamefont {Sayrin},
  \citenamefont {Bernu}, \citenamefont {Brune}, \citenamefont {Raimond},\ and\
  \citenamefont {Haroche}}]{deleglise_2008}%
  \BibitemOpen
  \bibfield  {author} {\bibinfo {author} {\bibfnamefont {Samuel}\ \bibnamefont
  {Del{\'e}glise}}, \bibinfo {author} {\bibfnamefont {Igor}\ \bibnamefont
  {Dotsenko}}, \bibinfo {author} {\bibfnamefont {Cl{\'e}ment}\ \bibnamefont
  {Sayrin}}, \bibinfo {author} {\bibfnamefont {Julien}\ \bibnamefont {Bernu}},
  \bibinfo {author} {\bibfnamefont {Michel}\ \bibnamefont {Brune}}, \bibinfo
  {author} {\bibfnamefont {Jean-Michel}\ \bibnamefont {Raimond}}, \ and\
  \bibinfo {author} {\bibfnamefont {Serge}\ \bibnamefont {Haroche}},\
  }\bibfield  {title} {\enquote {\bibinfo {title} {Reconstruction of
  non-classical cavity field states with snapshots of their decoherence},}\
  }\href {https://doi.org/10.1038/nature07288} {\bibfield  {journal} {\bibinfo
  {journal} {Nature}\ }\textbf {\bibinfo {volume} {455}},\ \bibinfo {pages}
  {510 EP --} (\bibinfo {year} {2008})}\BibitemShut {NoStop}%
\bibitem [{\citenamefont {Kiesel}\ and\ \citenamefont
  {Vogel}(2010)}]{kiesel_2010}%
  \BibitemOpen
  \bibfield  {author} {\bibinfo {author} {\bibfnamefont {T.}~\bibnamefont
  {Kiesel}}\ and\ \bibinfo {author} {\bibfnamefont {W.}~\bibnamefont {Vogel}},\
  }\bibfield  {title} {\enquote {\bibinfo {title} {Nonclassicality filters and
  quasiprobabilities},}\ }\href {\doibase 10.1103/PhysRevA.82.032107}
  {\bibfield  {journal} {\bibinfo  {journal} {Phys. Rev. A}\ }\textbf {\bibinfo
  {volume} {82}},\ \bibinfo {pages} {032107} (\bibinfo {year}
  {2010})}\BibitemShut {NoStop}%
\bibitem [{\citenamefont {Agudelo}\ \emph {et~al.}(2013)\citenamefont
  {Agudelo}, \citenamefont {Sperling},\ and\ \citenamefont
  {Vogel}}]{agudelo_2013}%
  \BibitemOpen
  \bibfield  {author} {\bibinfo {author} {\bibfnamefont {E.}~\bibnamefont
  {Agudelo}}, \bibinfo {author} {\bibfnamefont {J.}~\bibnamefont {Sperling}}, \
  and\ \bibinfo {author} {\bibfnamefont {W.}~\bibnamefont {Vogel}},\ }\bibfield
   {title} {\enquote {\bibinfo {title} {Quasiprobabilities for multipartite
  quantum correlations of light},}\ }\href {\doibase
  10.1103/PhysRevA.87.033811} {\bibfield  {journal} {\bibinfo  {journal} {Phys.
  Rev. A}\ }\textbf {\bibinfo {volume} {87}},\ \bibinfo {pages} {033811}
  (\bibinfo {year} {2013})}\BibitemShut {NoStop}%
\bibitem [{\citenamefont {Bohmann}\ and\ \citenamefont
  {Agudelo}(2020)}]{bohmann_2020}%
  \BibitemOpen
  \bibfield  {author} {\bibinfo {author} {\bibfnamefont {Martin}\ \bibnamefont
  {Bohmann}}\ and\ \bibinfo {author} {\bibfnamefont {Elizabeth}\ \bibnamefont
  {Agudelo}},\ }\bibfield  {title} {\enquote {\bibinfo {title} {Phase-space
  inequalities beyond negativities},}\ }\href {\doibase
  10.1103/PhysRevLett.124.133601} {\bibfield  {journal} {\bibinfo  {journal}
  {Phys. Rev. Lett.}\ }\textbf {\bibinfo {volume} {124}},\ \bibinfo {pages}
  {133601} (\bibinfo {year} {2020})}\BibitemShut {NoStop}%
\bibitem [{\citenamefont {Bohmann}\ \emph {et~al.}(2020)\citenamefont
  {Bohmann}, \citenamefont {Agudelo},\ and\ \citenamefont
  {Sperling}}]{bohmann_2020b}%
  \BibitemOpen
  \bibfield  {author} {\bibinfo {author} {\bibfnamefont {Martin}\ \bibnamefont
  {Bohmann}}, \bibinfo {author} {\bibfnamefont {Elizabeth}\ \bibnamefont
  {Agudelo}}, \ and\ \bibinfo {author} {\bibfnamefont {Jan}\ \bibnamefont
  {Sperling}},\ }\bibfield  {title} {\enquote {\bibinfo {title} {Probing
  nonclassicality with matrices of phase-space distributions},}\ }\href
  {\doibase 10.22331/q-2020-10-15-343} {\bibfield  {journal} {\bibinfo
  {journal} {{Quantum}}\ }\textbf {\bibinfo {volume} {4}},\ \bibinfo {pages}
  {343} (\bibinfo {year} {2020})}\BibitemShut {NoStop}%
\bibitem [{\citenamefont {Biagi}\ \emph {et~al.}(2021)\citenamefont {Biagi},
  \citenamefont {Bohmann}, \citenamefont {Agudelo}, \citenamefont {Bellini},\
  and\ \citenamefont {Zavatta}}]{biagi_2020}%
  \BibitemOpen
  \bibfield  {author} {\bibinfo {author} {\bibfnamefont {Nicola}\ \bibnamefont
  {Biagi}}, \bibinfo {author} {\bibfnamefont {Martin}\ \bibnamefont {Bohmann}},
  \bibinfo {author} {\bibfnamefont {Elizabeth}\ \bibnamefont {Agudelo}},
  \bibinfo {author} {\bibfnamefont {Marco}\ \bibnamefont {Bellini}}, \ and\
  \bibinfo {author} {\bibfnamefont {Alessandro}\ \bibnamefont {Zavatta}},\
  }\bibfield  {title} {\enquote {\bibinfo {title} {Experimental certification
  of nonclassicality via phase-space inequalities},}\ }\href {\doibase
  10.1103/PhysRevLett.126.023605} {\bibfield  {journal} {\bibinfo  {journal}
  {Phys. Rev. Lett.}\ }\textbf {\bibinfo {volume} {126}},\ \bibinfo {pages}
  {023605} (\bibinfo {year} {2021})}\BibitemShut {NoStop}%
\bibitem [{\citenamefont {Mari}\ \emph {et~al.}(2011)\citenamefont {Mari},
  \citenamefont {Kieling}, \citenamefont {Nielsen}, \citenamefont {Polzik},\
  and\ \citenamefont {Eisert}}]{mari_2011}%
  \BibitemOpen
  \bibfield  {author} {\bibinfo {author} {\bibfnamefont {A.}~\bibnamefont
  {Mari}}, \bibinfo {author} {\bibfnamefont {K.}~\bibnamefont {Kieling}},
  \bibinfo {author} {\bibfnamefont {B.~M.}\ \bibnamefont {Nielsen}},
  \bibinfo {author} {\bibfnamefont {E.~S.}\ \bibnamefont {Polzik}}, \ and\
  \bibinfo {author} {\bibfnamefont {J.}~\bibnamefont {Eisert}},\ }\bibfield
  {title} {\enquote {\bibinfo {title} {Directly estimating nonclassicality},}\
  }\href {\doibase 10.1103/PhysRevLett.106.010403} {\bibfield  {journal}
  {\bibinfo  {journal} {Phys. Rev. Lett.}\ }\textbf {\bibinfo {volume} {106}},\
  \bibinfo {pages} {010403} (\bibinfo {year} {2011})}\BibitemShut {NoStop}%
\bibitem [{\citenamefont {Nielsen}(2015)}]{nielsen_2015}%
  \BibitemOpen
  \bibfield  {author} {\bibinfo {author} {\bibfnamefont {Michael~A}\
  \bibnamefont {Nielsen}},\ }\href@noop {} {\emph {\bibinfo {title} {Neural
  networks and deep learning}}}\ (\bibinfo  {publisher} {Determination Press,
  San Francisco},\ \bibinfo {year} {2015})\BibitemShut {NoStop}%
\bibitem [{\citenamefont {Hentschel}\ and\ \citenamefont
  {Sanders}(2010)}]{hentschel_2010}%
  \BibitemOpen
  \bibfield  {author} {\bibinfo {author} {\bibfnamefont {Alexander}\
  \bibnamefont {Hentschel}}\ and\ \bibinfo {author} {\bibfnamefont {Barry~C.}\
  \bibnamefont {Sanders}},\ }\bibfield  {title} {\enquote {\bibinfo {title}
  {Machine learning for precise quantum measurement},}\ }\href {\doibase
  10.1103/PhysRevLett.104.063603} {\bibfield  {journal} {\bibinfo  {journal}
  {Phys. Rev. Lett.}\ }\textbf {\bibinfo {volume} {104}},\ \bibinfo {pages}
  {063603} (\bibinfo {year} {2010})}\BibitemShut {NoStop}%
\bibitem [{\citenamefont {Wiebe}\ \emph {et~al.}(2014)\citenamefont {Wiebe},
  \citenamefont {Granade}, \citenamefont {Ferrie},\ and\ \citenamefont
  {Cory}}]{wiebe_2014}%
  \BibitemOpen
  \bibfield  {author} {\bibinfo {author} {\bibfnamefont {Nathan}\ \bibnamefont
  {Wiebe}}, \bibinfo {author} {\bibfnamefont {Christopher}\ \bibnamefont
  {Granade}}, \bibinfo {author} {\bibfnamefont {Christopher}\ \bibnamefont
  {Ferrie}}, \ and\ \bibinfo {author} {\bibfnamefont {D.~G.}\ \bibnamefont
  {Cory}},\ }\bibfield  {title} {\enquote {\bibinfo {title} {Hamiltonian
  learning and certification using quantum resources},}\ }\href
  {http://dx.doi.org/10.1103/PhysRevLett.112.190501} {\bibfield  {journal}
  {\bibinfo  {journal} {Phys. Rev. Lett.}\ }\textbf {\bibinfo {volume} {112}},\
  \bibinfo {pages} {190501} (\bibinfo {year} {2014})}\BibitemShut {NoStop}%
\bibitem [{\citenamefont {Magesan}\ \emph {et~al.}(2015)\citenamefont
  {Magesan}, \citenamefont {Gambetta}, \citenamefont {C\'orcoles},\ and\
  \citenamefont {Chow}}]{magesan_2015}%
  \BibitemOpen
  \bibfield  {author} {\bibinfo {author} {\bibfnamefont {Easwar}\ \bibnamefont
  {Magesan}}, \bibinfo {author} {\bibfnamefont {Jay~M.}\ \bibnamefont
  {Gambetta}}, \bibinfo {author} {\bibfnamefont {A.~D.}\ \bibnamefont
  {C\'orcoles}}, \ and\ \bibinfo {author} {\bibfnamefont {Jerry~M.}\
  \bibnamefont {Chow}},\ }\bibfield  {title} {\enquote {\bibinfo {title}
  {Machine learning for discriminating quantum measurement trajectories and
  improving readout},}\ }\href {\doibase 10.1103/PhysRevLett.114.200501}
  {\bibfield  {journal} {\bibinfo  {journal} {Phys. Rev. Lett.}\ }\textbf
  {\bibinfo {volume} {114}},\ \bibinfo {pages} {200501} (\bibinfo {year}
  {2015})}\BibitemShut {NoStop}%
\bibitem [{\citenamefont {Krenn}\ \emph {et~al.}(2016)\citenamefont {Krenn},
  \citenamefont {Malik}, \citenamefont {Fickler}, \citenamefont {Lapkiewicz},\
  and\ \citenamefont {Zeilinger}}]{krenn_2016}%
  \BibitemOpen
  \bibfield  {author} {\bibinfo {author} {\bibfnamefont {Mario}\ \bibnamefont
  {Krenn}}, \bibinfo {author} {\bibfnamefont {Mehul}\ \bibnamefont {Malik}},
  \bibinfo {author} {\bibfnamefont {Robert}\ \bibnamefont {Fickler}}, \bibinfo
  {author} {\bibfnamefont {Radek}\ \bibnamefont {Lapkiewicz}}, \ and\ \bibinfo
  {author} {\bibfnamefont {Anton}\ \bibnamefont {Zeilinger}},\ }\bibfield
  {title} {\enquote {\bibinfo {title} {Automated search for new quantum
  experiments},}\ }\href {\doibase 10.1103/PhysRevLett.116.090405} {\bibfield
  {journal} {\bibinfo  {journal} {Phys. Rev. Lett.}\ }\textbf {\bibinfo
  {volume} {116}},\ \bibinfo {pages} {090405} (\bibinfo {year}
  {2016})}\BibitemShut {NoStop}%
\bibitem [{\citenamefont {van Nieuwenburg}\ \emph {et~al.}(2017)\citenamefont
  {van Nieuwenburg}, \citenamefont {Liu},\ and\ \citenamefont
  {Huber}}]{van_nieuwenburg_2017}%
  \BibitemOpen
  \bibfield  {author} {\bibinfo {author} {\bibfnamefont {Evert P. L.}\
  \bibnamefont {van Nieuwenburg}}, \bibinfo {author} {\bibfnamefont {Ye-Hua}\
  \bibnamefont {Liu}}, \ and\ \bibinfo {author} {\bibfnamefont {Sebastian D.}\
  \bibnamefont {Huber}},\ }\bibfield  {title} {\enquote {\bibinfo {title}
  {Learning phase transitions by confusion},}\ }\href {\doibase
  10.1038/nphys4037} {\bibfield  {journal} {\bibinfo  {journal} {Nat. Phys.}\
  }\textbf {\bibinfo {volume} {13}},\ \bibinfo {pages} {435–439} (\bibinfo
  {year} {2017})}\BibitemShut {NoStop}%
\bibitem [{\citenamefont {Carleo}\ and\ \citenamefont
  {Troyer}(2017)}]{carleo_2017}%
  \BibitemOpen
  \bibfield  {author} {\bibinfo {author} {\bibfnamefont {Giuseppe}\
  \bibnamefont {Carleo}}\ and\ \bibinfo {author} {\bibfnamefont {Matthias}\
  \bibnamefont {Troyer}},\ }\bibfield  {title} {\enquote {\bibinfo {title}
  {Solving the quantum many-body problem with artificial neural networks},}\
  }\href {\doibase 10.1126/science.aag2302} {\bibfield  {journal} {\bibinfo
  {journal} {Science}\ }\textbf {\bibinfo {volume} {355}},\ \bibinfo {pages}
  {602--606} (\bibinfo {year} {2017})}\BibitemShut {NoStop}%
\bibitem [{\citenamefont {Torlai}\ \emph {et~al.}(2018)\citenamefont {Torlai},
  \citenamefont {Mazzola}, \citenamefont {Carrasquilla}, \citenamefont
  {Troyer}, \citenamefont {Melko},\ and\ \citenamefont {Carleo}}]{torlai_2018}%
  \BibitemOpen
  \bibfield  {author} {\bibinfo {author} {\bibfnamefont {Giacomo}\ \bibnamefont
  {Torlai}}, \bibinfo {author} {\bibfnamefont {Guglielmo}\ \bibnamefont
  {Mazzola}}, \bibinfo {author} {\bibfnamefont {Juan}\ \bibnamefont
  {Carrasquilla}}, \bibinfo {author} {\bibfnamefont {Matthias}\ \bibnamefont
  {Troyer}}, \bibinfo {author} {\bibfnamefont {Roger}\ \bibnamefont {Melko}}, \
  and\ \bibinfo {author} {\bibfnamefont {Giuseppe}\ \bibnamefont {Carleo}},\
  }\bibfield  {title} {\enquote {\bibinfo {title} {Neural-network quantum state
  tomography},}\ }\href {\doibase 10.1038/s41567-018-0048-5} {\bibfield
  {journal} {\bibinfo  {journal} {Nat. Phys.}\ }\textbf {\bibinfo {volume}
  {14}},\ \bibinfo {pages} {447–450} (\bibinfo {year} {2018})}\BibitemShut
  {NoStop}%
\bibitem [{\citenamefont {Lumino}\ \emph {et~al.}(2018)\citenamefont {Lumino},
  \citenamefont {Polino}, \citenamefont {Rab}, \citenamefont {Milani},
  \citenamefont {Spagnolo}, \citenamefont {Wiebe},\ and\ \citenamefont
  {Sciarrino}}]{lumino_2018}%
  \BibitemOpen
  \bibfield  {author} {\bibinfo {author} {\bibfnamefont {Alessandro}\
  \bibnamefont {Lumino}}, \bibinfo {author} {\bibfnamefont {Emanuele}\
  \bibnamefont {Polino}}, \bibinfo {author} {\bibfnamefont {Adil~S.}\
  \bibnamefont {Rab}}, \bibinfo {author} {\bibfnamefont {Giorgio}\ \bibnamefont
  {Milani}}, \bibinfo {author} {\bibfnamefont {Nicol\`o}\ \bibnamefont
  {Spagnolo}}, \bibinfo {author} {\bibfnamefont {Nathan}\ \bibnamefont
  {Wiebe}}, \ and\ \bibinfo {author} {\bibfnamefont {Fabio}\ \bibnamefont
  {Sciarrino}},\ }\bibfield  {title} {\enquote {\bibinfo {title} {Experimental
  phase estimation enhanced by machine learning},}\ }\href {\doibase
  10.1103/PhysRevApplied.10.044033} {\bibfield  {journal} {\bibinfo  {journal}
  {Phys. Rev. Applied}\ }\textbf {\bibinfo {volume} {10}},\ \bibinfo {pages}
  {044033} (\bibinfo {year} {2018})}\BibitemShut {NoStop}%
\bibitem [{\citenamefont {Bukov}\ \emph {et~al.}(2018)\citenamefont {Bukov},
  \citenamefont {Day}, \citenamefont {Sels}, \citenamefont {Weinberg},
  \citenamefont {Polkovnikov},\ and\ \citenamefont {Mehta}}]{bukov_2018}%
  \BibitemOpen
  \bibfield  {author} {\bibinfo {author} {\bibfnamefont {Marin}\ \bibnamefont
  {Bukov}}, \bibinfo {author} {\bibfnamefont {Alexandre G.~R.}\ \bibnamefont
  {Day}}, \bibinfo {author} {\bibfnamefont {Dries}\ \bibnamefont {Sels}},
  \bibinfo {author} {\bibfnamefont {Phillip}\ \bibnamefont {Weinberg}},
  \bibinfo {author} {\bibfnamefont {Anatoli}\ \bibnamefont {Polkovnikov}}, \
  and\ \bibinfo {author} {\bibfnamefont {Pankaj}\ \bibnamefont {Mehta}},\
  }\bibfield  {title} {\enquote {\bibinfo {title} {Reinforcement learning in
  different phases of quantum control},}\ }\href {\doibase
  10.1103/PhysRevX.8.031086} {\bibfield  {journal} {\bibinfo  {journal} {Phys.
  Rev. X}\ }\textbf {\bibinfo {volume} {8}},\ \bibinfo {pages} {031086}
  (\bibinfo {year} {2018})}\BibitemShut {NoStop}%
\bibitem [{\citenamefont {F\"osel}\ \emph {et~al.}(2018)\citenamefont
  {F\"osel}, \citenamefont {Tighineanu}, \citenamefont {Weiss},\ and\
  \citenamefont {Marquardt}}]{fosel_2018}%
  \BibitemOpen
  \bibfield  {author} {\bibinfo {author} {\bibfnamefont {Thomas}\ \bibnamefont
  {F\"osel}}, \bibinfo {author} {\bibfnamefont {Petru}\ \bibnamefont
  {Tighineanu}}, \bibinfo {author} {\bibfnamefont {Talitha}\ \bibnamefont
  {Weiss}}, \ and\ \bibinfo {author} {\bibfnamefont {Florian}\ \bibnamefont
  {Marquardt}},\ }\bibfield  {title} {\enquote {\bibinfo {title} {Reinforcement
  learning with neural networks for quantum feedback},}\ }\href {\doibase
  10.1103/PhysRevX.8.031084} {\bibfield  {journal} {\bibinfo  {journal} {Phys.
  Rev. X}\ }\textbf {\bibinfo {volume} {8}},\ \bibinfo {pages} {031084}
  (\bibinfo {year} {2018})}\BibitemShut {NoStop}%
\bibitem [{\citenamefont {Canabarro}\ \emph {et~al.}(2019)\citenamefont
  {Canabarro}, \citenamefont {Brito},\ and\ \citenamefont
  {Chaves}}]{canabarro_2019}%
  \BibitemOpen
  \bibfield  {author} {\bibinfo {author} {\bibfnamefont {Askery}\ \bibnamefont
  {Canabarro}}, \bibinfo {author} {\bibfnamefont {Samura\'{\i}}\ \bibnamefont
  {Brito}}, \ and\ \bibinfo {author} {\bibfnamefont {Rafael}\ \bibnamefont
  {Chaves}},\ }\bibfield  {title} {\enquote {\bibinfo {title} {{Machine
  Learning Nonlocal Correlations}},}\ }\href {\doibase
  10.1103/PhysRevLett.122.200401} {\bibfield  {journal} {\bibinfo  {journal}
  {Phys. Rev. Lett.}\ }\textbf {\bibinfo {volume} {122}},\ \bibinfo {pages}
  {200401} (\bibinfo {year} {2019})}\BibitemShut {NoStop}%
\bibitem [{\citenamefont {Cimini}\ \emph {et~al.}(2019)\citenamefont {Cimini},
  \citenamefont {Gianani}, \citenamefont {Spagnolo}, \citenamefont {Leccese},
  \citenamefont {Sciarrino},\ and\ \citenamefont {Barbieri}}]{cimini_2019}%
  \BibitemOpen
  \bibfield  {author} {\bibinfo {author} {\bibfnamefont {Valeria}\ \bibnamefont
  {Cimini}}, \bibinfo {author} {\bibfnamefont {Ilaria}\ \bibnamefont
  {Gianani}}, \bibinfo {author} {\bibfnamefont {Nicol\`o}\ \bibnamefont
  {Spagnolo}}, \bibinfo {author} {\bibfnamefont {Fabio}\ \bibnamefont
  {Leccese}}, \bibinfo {author} {\bibfnamefont {Fabio}\ \bibnamefont
  {Sciarrino}}, \ and\ \bibinfo {author} {\bibfnamefont {Marco}\ \bibnamefont
  {Barbieri}},\ }\bibfield  {title} {\enquote {\bibinfo {title} {Calibration of
  quantum sensors by neural networks},}\ }\href {\doibase
  10.1103/PhysRevLett.123.230502} {\bibfield  {journal} {\bibinfo  {journal}
  {Phys. Rev. Lett.}\ }\textbf {\bibinfo {volume} {123}},\ \bibinfo {pages}
  {230502} (\bibinfo {year} {2019})}\BibitemShut {NoStop}%
\bibitem [{\citenamefont {Poulsen~Nautrup}\ \emph {et~al.}(2019)\citenamefont
  {Poulsen~Nautrup}, \citenamefont {Delfosse}, \citenamefont {Dunjko},
  \citenamefont {Briegel},\ and\ \citenamefont {Friis}}]{nautrup_2019}%
  \BibitemOpen
  \bibfield  {author} {\bibinfo {author} {\bibfnamefont {Hendrik}\ \bibnamefont
  {Poulsen~Nautrup}}, \bibinfo {author} {\bibfnamefont {Nicolas}\ \bibnamefont
  {Delfosse}}, \bibinfo {author} {\bibfnamefont {Vedran}\ \bibnamefont
  {Dunjko}}, \bibinfo {author} {\bibfnamefont {Hans~J.}\ \bibnamefont
  {Briegel}}, \ and\ \bibinfo {author} {\bibfnamefont {Nicolai}\ \bibnamefont
  {Friis}},\ }\bibfield  {title} {\enquote {\bibinfo {title} {Optimizing
  {Q}uantum {E}rror {C}orrection {C}odes with {R}einforcement {L}earning},}\
  }\href {\doibase 10.22331/q-2019-12-16-215} {\bibfield  {journal} {\bibinfo
  {journal} {{Quantum}}\ }\textbf {\bibinfo {volume} {3}},\ \bibinfo {pages}
  {215} (\bibinfo {year} {2019})}\BibitemShut {NoStop}%
\bibitem [{\citenamefont {Agresti}\ \emph {et~al.}(2019)\citenamefont
  {Agresti}, \citenamefont {Viggianiello}, \citenamefont {Flamini},
  \citenamefont {Spagnolo}, \citenamefont {Crespi}, \citenamefont {Osellame},
  \citenamefont {Wiebe},\ and\ \citenamefont {Sciarrino}}]{agresti_2019}%
  \BibitemOpen
  \bibfield  {author} {\bibinfo {author} {\bibfnamefont {Iris}\ \bibnamefont
  {Agresti}}, \bibinfo {author} {\bibfnamefont {Niko}\ \bibnamefont
  {Viggianiello}}, \bibinfo {author} {\bibfnamefont {Fulvio}\ \bibnamefont
  {Flamini}}, \bibinfo {author} {\bibfnamefont {Nicol\`o}\ \bibnamefont
  {Spagnolo}}, \bibinfo {author} {\bibfnamefont {Andrea}\ \bibnamefont
  {Crespi}}, \bibinfo {author} {\bibfnamefont {Roberto}\ \bibnamefont
  {Osellame}}, \bibinfo {author} {\bibfnamefont {Nathan}\ \bibnamefont
  {Wiebe}}, \ and\ \bibinfo {author} {\bibfnamefont {Fabio}\ \bibnamefont
  {Sciarrino}},\ }\bibfield  {title} {\enquote {\bibinfo {title} {Pattern
  recognition techniques for boson sampling validation},}\ }\href {\doibase
  10.1103/PhysRevX.9.011013} {\bibfield  {journal} {\bibinfo  {journal} {Phys.
  Rev. X}\ }\textbf {\bibinfo {volume} {9}},\ \bibinfo {pages} {011013}
  (\bibinfo {year} {2019})}\BibitemShut {NoStop}%
\bibitem [{\citenamefont {Gebhart}\ and\ \citenamefont
  {Bohmann}(2020)}]{gebhart_2020}%
  \BibitemOpen
  \bibfield  {author} {\bibinfo {author} {\bibfnamefont {Valentin}\
  \bibnamefont {Gebhart}}\ and\ \bibinfo {author} {\bibfnamefont {Martin}\
  \bibnamefont {Bohmann}},\ }\bibfield  {title} {\enquote {\bibinfo {title}
  {Neural-network approach for identifying nonclassicality from click-counting
  data},}\ }\href {\doibase 10.1103/PhysRevResearch.2.023150} {\bibfield
  {journal} {\bibinfo  {journal} {Phys. Rev. Research}\ }\textbf {\bibinfo
  {volume} {2}},\ \bibinfo {pages} {023150} (\bibinfo {year}
  {2020})}\BibitemShut {NoStop}%
\bibitem [{\citenamefont {Cimini}\ \emph {et~al.}(2020)\citenamefont {Cimini},
  \citenamefont {Barbieri}, \citenamefont {Treps}, \citenamefont {Walschaers},\
  and\ \citenamefont {Parigi}}]{cimini_2020}%
  \BibitemOpen
  \bibfield  {author} {\bibinfo {author} {\bibfnamefont {Valeria}\ \bibnamefont
  {Cimini}}, \bibinfo {author} {\bibfnamefont {Marco}\ \bibnamefont
  {Barbieri}}, \bibinfo {author} {\bibfnamefont {Nicolas}\ \bibnamefont
  {Treps}}, \bibinfo {author} {\bibfnamefont {Mattia}\ \bibnamefont
  {Walschaers}}, \ and\ \bibinfo {author} {\bibfnamefont {Valentina}\
  \bibnamefont {Parigi}},\ }\bibfield  {title} {\enquote {\bibinfo {title}
  {Neural networks for detecting multimode wigner negativity},}\ }\href
  {\doibase 10.1103/PhysRevLett.125.160504} {\bibfield  {journal} {\bibinfo
  {journal} {Phys. Rev. Lett.}\ }\textbf {\bibinfo {volume} {125}},\ \bibinfo
  {pages} {160504} (\bibinfo {year} {2020})}\BibitemShut {NoStop}%
\bibitem [{\citenamefont {Tiunov}\ \emph {et~al.}(2020)\citenamefont {Tiunov},
  \citenamefont {(Vyborova)}, \citenamefont {Ulanov}, \citenamefont {Lvovsky},\
  and\ \citenamefont {Fedorov}}]{tiunov_2020}%
  \BibitemOpen
  \bibfield  {author} {\bibinfo {author} {\bibfnamefont {E.~S.}\ \bibnamefont
  {Tiunov}}, \bibinfo {author} {\bibfnamefont {V.~V.~Tiunova}\ \bibnamefont
  {(Vyborova)}}, \bibinfo {author} {\bibfnamefont {A.~E.}\ \bibnamefont
  {Ulanov}}, \bibinfo {author} {\bibfnamefont {A.~I.}\ \bibnamefont {Lvovsky}},
  \ and\ \bibinfo {author} {\bibfnamefont {A.~K.}\ \bibnamefont {Fedorov}},\
  }\bibfield  {title} {\enquote {\bibinfo {title} {Experimental quantum
  homodyne tomography via machine learning},}\ }\href {\doibase
  10.1364/OPTICA.389482} {\bibfield  {journal} {\bibinfo  {journal} {Optica}\
  }\textbf {\bibinfo {volume} {7}},\ \bibinfo {pages} {448--454} (\bibinfo
  {year} {2020})}\BibitemShut {NoStop}%
\bibitem [{\citenamefont {Nolan}\ \emph {et~al.}(2020)\citenamefont {Nolan},
  \citenamefont {Smerzi},\ and\ \citenamefont {Pezz\`{e}}}]{nolan_2020}%
  \BibitemOpen
  \bibfield  {author} {\bibinfo {author} {\bibfnamefont {Samuel~P.}\
  \bibnamefont {Nolan}}, \bibinfo {author} {\bibfnamefont {Augusto}\
  \bibnamefont {Smerzi}}, \ and\ \bibinfo {author} {\bibfnamefont {Luca}\
  \bibnamefont {Pezz\`{e}}},\ }\href@noop {} {\enquote {\bibinfo {title} {A
  machine learning approach to bayesian parameter estimation},}\ } (\bibinfo
  {year} {2020}),\ \Eprint {http://arxiv.org/abs/arXiv:2006.02369}
  {arXiv:2006.02369} \BibitemShut {NoStop}%
\bibitem [{\citenamefont {Dunjko}\ and\ \citenamefont
  {Briegel}(2018)}]{dunjko_2018}%
  \BibitemOpen
  \bibfield  {author} {\bibinfo {author} {\bibfnamefont {Vedran}\ \bibnamefont
  {Dunjko}}\ and\ \bibinfo {author} {\bibfnamefont {Hans~J}\ \bibnamefont
  {Briegel}},\ }\bibfield  {title} {\enquote {\bibinfo {title} {Machine
  learning {\&} artificial intelligence in the quantum domain: a review of
  recent progress},}\ }\href {\doibase 10.1088/1361-6633/aab406} {\bibfield
  {journal} {\bibinfo  {journal} {Rep. Prog. Phys.}\ }\textbf {\bibinfo
  {volume} {81}},\ \bibinfo {pages} {074001} (\bibinfo {year}
  {2018})}\BibitemShut {NoStop}%
\bibitem [{\citenamefont {Ahmed}\ \emph {et~al.}(2020)\citenamefont {Ahmed},
  \citenamefont {Mu{\~n}oz}, \citenamefont {Nori},\ and\ \citenamefont
  {Kockum}}]{ahmed_2020}%
  \BibitemOpen
  \bibfield  {author} {\bibinfo {author} {\bibfnamefont {S.}~\bibnamefont
  {Ahmed}}, \bibinfo {author} {\bibfnamefont {C.~S.}\ \bibnamefont
  {Mu{\~n}oz}}, \bibinfo {author} {\bibfnamefont {F.}~\bibnamefont {Nori}}, \
  and\ \bibinfo {author} {\bibfnamefont {A.~F.}\ \bibnamefont {Kockum}},\
  }\href@noop {} {\enquote {\bibinfo {title} {Classification and reconstruction
  of optical quantum states with deep neural networks},}\ } (\bibinfo {year}
  {2020}),\ \Eprint {http://arxiv.org/abs/arXiv:2012.02185} {arXiv:2012.02185}
  \BibitemShut {NoStop}%
\bibitem [{\citenamefont {Carmichael}(1987)}]{Carmichael_1987}%
  \BibitemOpen
  \bibfield  {author} {\bibinfo {author} {\bibfnamefont {H.~J.}\ \bibnamefont
  {Carmichael}},\ }\bibfield  {title} {\enquote {\bibinfo {title} {Spectrum of
  squeezing and photocurrent shot noise: a normally ordered treatment},}\
  }\href {\doibase 10.1364/JOSAB.4.001588} {\bibfield  {journal} {\bibinfo
  {journal} {J. Opt. Soc. Am. B}\ }\textbf {\bibinfo {volume} {4}},\ \bibinfo
  {pages} {1588--1603} (\bibinfo {year} {1987})}\BibitemShut {NoStop}%
\bibitem [{\citenamefont {Braunstein}(1990)}]{Braunstein_1990}%
  \BibitemOpen
  \bibfield  {author} {\bibinfo {author} {\bibfnamefont {Samuel~L.}\
  \bibnamefont {Braunstein}},\ }\bibfield  {title} {\enquote {\bibinfo {title}
  {Homodyne statistics},}\ }\href {\doibase 10.1103/PhysRevA.42.474} {\bibfield
   {journal} {\bibinfo  {journal} {Phys. Rev. A}\ }\textbf {\bibinfo {volume}
  {42}},\ \bibinfo {pages} {474--481} (\bibinfo {year} {1990})}\BibitemShut
  {NoStop}%
\bibitem [{\citenamefont {Vogel}\ and\ \citenamefont
  {Grabow}(1993)}]{Vogel_1993}%
  \BibitemOpen
  \bibfield  {author} {\bibinfo {author} {\bibfnamefont {Werner}\ \bibnamefont
  {Vogel}}\ and\ \bibinfo {author} {\bibfnamefont {Jens}\ \bibnamefont
  {Grabow}},\ }\bibfield  {title} {\enquote {\bibinfo {title} {Statistics of
  difference events in homodyne detection},}\ }\href {\doibase
  10.1103/PhysRevA.47.4227} {\bibfield  {journal} {\bibinfo  {journal} {Phys.
  Rev. A}\ }\textbf {\bibinfo {volume} {47}},\ \bibinfo {pages} {4227--4235}
  (\bibinfo {year} {1993})}\BibitemShut {NoStop}%
\bibitem [{Not()}]{Note1}%
  \BibitemOpen
  \href@noop {} {}\bibinfo {note} {To preserve essential information
  with this restriction of detection values, we only simulate states for which
  the probability of producing an event $\left|x\right|>8$ is small
  ($<10^{-6}$), see Appendix. If the network is to be applied to states which
  produce values outside of this domain, the grid has to be
  adjusted.}\BibitemShut {Stop}%
\bibitem [{\citenamefont {Agudelo}\ \emph {et~al.}(2015)\citenamefont
  {Agudelo}, \citenamefont {Sperling}, \citenamefont {Vogel}, \citenamefont
  {K\"ohnke}, \citenamefont {Mraz},\ and\ \citenamefont {Hage}}]{agudelo_2015}%
  \BibitemOpen
  \bibfield  {author} {\bibinfo {author} {\bibfnamefont {E.}~\bibnamefont
  {Agudelo}}, \bibinfo {author} {\bibfnamefont {J.}~\bibnamefont {Sperling}},
  \bibinfo {author} {\bibfnamefont {W.}~\bibnamefont {Vogel}}, \bibinfo
  {author} {\bibfnamefont {S.}~\bibnamefont {K\"ohnke}}, \bibinfo {author}
  {\bibfnamefont {M.}~\bibnamefont {Mraz}}, \ and\ \bibinfo {author}
  {\bibfnamefont {B.}~\bibnamefont {Hage}},\ }\bibfield  {title} {\enquote
  {\bibinfo {title} {Continuous sampling of the squeezed-state
  nonclassicality},}\ }\href {\doibase 10.1103/PhysRevA.92.033837} {\bibfield
  {journal} {\bibinfo  {journal} {Phys. Rev. A}\ }\textbf {\bibinfo {volume}
  {92}},\ \bibinfo {pages} {033837} (\bibinfo {year} {2015})}\BibitemShut
  {NoStop}%
\bibitem [{\citenamefont {Vogel}\ and\ \citenamefont
  {Welsch}(2006)}]{vogel_2006}%
  \BibitemOpen
  \bibfield  {author} {\bibinfo {author} {\bibfnamefont {W.}~\bibnamefont
  {Vogel}}\ and\ \bibinfo {author} {\bibfnamefont {D.G.}\ \bibnamefont
  {Welsch}},\ }\href {https://books.google.it/books?id=GE7FuoEaGQAC} {\emph
  {\bibinfo {title} {Quantum Optics}}}\ (\bibinfo  {publisher} {Wiley},\
  \bibinfo {year} {2006})\BibitemShut {NoStop}%
\bibitem [{\citenamefont {Zavatta}\ \emph {et~al.}(2005)\citenamefont
  {Zavatta}, \citenamefont {Viciani},\ and\ \citenamefont
  {Bellini}}]{zavatta_2005}%
  \BibitemOpen
  \bibfield  {author} {\bibinfo {author} {\bibfnamefont {Alessandro}\
  \bibnamefont {Zavatta}}, \bibinfo {author} {\bibfnamefont {Silvia}\
  \bibnamefont {Viciani}}, \ and\ \bibinfo {author} {\bibfnamefont {Marco}\
  \bibnamefont {Bellini}},\ }\bibfield  {title} {\enquote {\bibinfo {title}
  {Single-photon excitation of a coherent state: Catching the elementary step
  of stimulated light emission},}\ }\href {\doibase 10.1103/PhysRevA.72.023820}
  {\bibfield  {journal} {\bibinfo  {journal} {Phys. Rev. A}\ }\textbf {\bibinfo
  {volume} {72}},\ \bibinfo {pages} {023820} (\bibinfo {year}
  {2005})}\BibitemShut {NoStop}%
\bibitem [{\citenamefont {Zavatta}\ \emph {et~al.}(2004)\citenamefont
  {Zavatta}, \citenamefont {Viciani},\ and\ \citenamefont
  {Bellini}}]{zavatta_2004}%
  \BibitemOpen
  \bibfield  {author} {\bibinfo {author} {\bibfnamefont {Alessandro}\
  \bibnamefont {Zavatta}}, \bibinfo {author} {\bibfnamefont {Silvia}\
  \bibnamefont {Viciani}}, \ and\ \bibinfo {author} {\bibfnamefont {Marco}\
  \bibnamefont {Bellini}},\ }\bibfield  {title} {\enquote {\bibinfo {title}
  {Quantum-to-classical transition with single-photon-added coherent states of
  light},}\ }\href {\doibase 10.1126/science.1103190} {\bibfield  {journal}
  {\bibinfo  {journal} {Science}\ }\textbf {\bibinfo {volume} {306}},\ \bibinfo
  {pages} {660--662} (\bibinfo {year} {2004})}\BibitemShut {NoStop}%
\bibitem [{\citenamefont {Filippov}\ \emph {et~al.}(2013)\citenamefont
  {Filippov}, \citenamefont {Man'ko}, \citenamefont {Coelho}, \citenamefont
  {Zavatta},\ and\ \citenamefont {Bellini}}]{filippov_2013}%
  \BibitemOpen
  \bibfield  {author} {\bibinfo {author} {\bibfnamefont {S~N}\ \bibnamefont
  {Filippov}}, \bibinfo {author} {\bibfnamefont {V~I}\ \bibnamefont {Man'ko}},
  \bibinfo {author} {\bibfnamefont {A~S}\ \bibnamefont {Coelho}}, \bibinfo
  {author} {\bibfnamefont {A}~\bibnamefont {Zavatta}}, \ and\ \bibinfo {author}
  {\bibfnamefont {M}~\bibnamefont {Bellini}},\ }\bibfield  {title} {\enquote
  {\bibinfo {title} {Single-photon-added coherent states: estimation of
  parameters and fidelity of the optical homodyne detection},}\ }\href
  {\doibase 10.1088/0031-8949/2013/t153/014025} {\bibfield  {journal} {\bibinfo
   {journal} {Physica Scripta}\ }\textbf {\bibinfo {volume} {T153}},\ \bibinfo
  {pages} {014025} (\bibinfo {year} {2013})}\BibitemShut {NoStop}%
\bibitem [{\citenamefont {Jeong}\ \emph {et~al.}(2014)\citenamefont {Jeong},
  \citenamefont {Zavatta}, \citenamefont {Kang}, \citenamefont {Lee},
  \citenamefont {Costanzo}, \citenamefont {Grandi}, \citenamefont {Ralph},\
  and\ \citenamefont {Bellini}}]{jeong_2014}%
  \BibitemOpen
  \bibfield  {author} {\bibinfo {author} {\bibfnamefont {Hyunseok}\
  \bibnamefont {Jeong}}, \bibinfo {author} {\bibfnamefont {Alessandro}\
  \bibnamefont {Zavatta}}, \bibinfo {author} {\bibfnamefont {Minsu}\
  \bibnamefont {Kang}}, \bibinfo {author} {\bibfnamefont {Seung-Woo}\
  \bibnamefont {Lee}}, \bibinfo {author} {\bibfnamefont {Luca~S.}\ \bibnamefont
  {Costanzo}}, \bibinfo {author} {\bibfnamefont {Samuele}\ \bibnamefont
  {Grandi}}, \bibinfo {author} {\bibfnamefont {Timothy~C.}\ \bibnamefont
  {Ralph}}, \ and\ \bibinfo {author} {\bibfnamefont {Marco}\ \bibnamefont
  {Bellini}},\ }\bibfield  {title} {\enquote {\bibinfo {title} {Generation of
  hybrid entanglement of light},}\ }\href
  {https://doi.org/10.1038/nphoton.2014.136} {\bibfield  {journal} {\bibinfo
  {journal} {Nat. Photon.}\ }\textbf {\bibinfo {volume} {8}},\ \bibinfo {pages}
  {564 EP --} (\bibinfo {year} {2014})}\BibitemShut {NoStop}%
\end{thebibliography}
\end{document}